\documentclass[prb,preprint,amsmath,amssymb]{revtex4}

\usepackage{color}
\begin{document}
\author{Kevin Leung$^{1,*}$ and Katherine L.~Jungjohann$^2$}
\affiliation{$^1$Sandia National Laboratories, 
Albuquerque, NM 87185, United States\\
$^2$Center for Integrated Nanotechnologies, Sandia National Laboratories,
Albuquerque, NM 87185, United States\\
\tt kleung@sandia.gov}
\date{\today}
\title{Spatial Heterogeneities and Onset of Passivation 
Breakdown at Lithium Anode Interfaces}

\input epsf
\renewcommand{\thetable}{\arabic{table}}

\begin{abstract}

Effective passivation of lithium metal surfaces, and prevention of
battery-shorting lithium dendrite growth, are critical for implementing
lithium-metal-anodes for batteries with increased power densities.  Nanoscale
surface heterogeneities can be ``hot spots'' where anode passivation breaks
down.  Motivated by the observation of lithium dendrites in pores and grain
boundaries in all-solid batteries, we examine lithium metal surfaces covered
with Li$_2$O and/or LiF thin films with grain boundaries in them.  Electronic
structure calculations show that, at $>$0.25~V computed equilibrium
overpotential, Li$_2$O grain boundaries with sufficiently large pores can
accommodate Li$^{(0)}$ atoms which aid $e^-$ leakage and passivation
breakdown.  Strain often accompanies Li-insertion; applying a $\sim$1.7\%
strain already lowers the computed overpotential to 0.1~V.  Lithium metal
nanostructures as thin as 12~\AA\, are thermodynamically favored inside
cracks in Li$_2$O films, becoming ``incipient lithium filaments.''  LiF films
are more resistant to lithium metal growth.  The models used herein should in
turn inform passivating strategies in all-solid-state batteries.

\end{abstract}

\maketitle

\section*{INTRODUCTION}
 
Lithium metal is the most gravimetrically efficient anode candidate material
for next-generation batteries.\cite{li1}  Replacing graphite with Li anode
would yield a 3$\times$ increase in anode capacity.  As Li(s) is extremely
electronegative, and reacts with almost all electrolytes in liquid
electrolyte-based lithium ion batteries (henceforth LELIB), it requires surface
passivation films to block electron tunneling to, and direct chemical contact
with the electrolyte.\cite{pnnl,cui,archer,jungjohann,dudney1} The innermost
layers of such protective films tend to be inorganic in nature.  They are
either formed naturally from electrolyte decomposition products (``solid
electrolyte interphase'' or ``SEI''), are artificial coatings, or are formed
with solid electrolytes.\cite{pearse,dudney}

Even when using protection schemes such as coating lithium metal with
passivation films, lithium dendrites may still grow
from lithium metal anodes under adverse (e.g., overpotential) conditions.
These dendrites can penetrate the separator containing the liquid electrolyte,
reach the cathode,\cite{naturetem,garcia} and cause a short circuit and possibly
a fire.  Dendrites are therefore significant battery reliability and safety
concerns.  Historically, many studies of dendrite formation in liquid
electrolyte-based lithium ion batteries (henceforth LELIB) with lithium metal
anodes have focused on homogeneous
films.\cite{newman,santosh1,holzwarth,yue1,yue}  One popular viewpoint is
adopted from electroplating of metal in water, where a solid blocking film is
absent, and dendrites are assume to arise from spontaneous, local fluctuations
of electric fields.  While this viewpoint leads to useful mitigating
strategies,\cite{pnnl} it ignores the fact that lithium nucleation and
dissolution occurs at
particular locations on the SEI film in LELIB;\cite{jungjohann} it cannot
describe the entire passivation breakdown mechanism. The influence of SEI
spatial heterogeneity has been addressed in transition electron
microscopy studies.\cite{jungjohann,naturetem}  However, most lithium dendrites
cannot be imaged in battery settings until they are at least 100~nm in diameter,
by which time their growth is rapid and unmanageable.  Atomic-scale
inhomogeneities, from which dendrites may originate, are difficult to image due
to the small length scale, low-scattering elemental composition, and the
buried nature of the interface.  Similarly, modeling
of dendrite growth dynamics has focused on the phase-field method,
which deals with meso-, not atomic, lengthscales.\cite{garcia}

Spatial inhomogeneities or ``hot spots'' on the surfaces of graphite anodes
and LiCoO$_2$ cathodes in LELIB are known to exist, leading to battery
failure.\cite{harris}  While solid electrolyte-based all-solid lithium ion
batteries (henceforth SELIB) are inherently safer than LELIB, lithium metal
also reacts with many Li$^+$-conducting solid electrolytes.  In particular,
dendrites are found to grow inside pores and grain boundaries in some solid
electrolyte materials.\cite{pore,pore1} These findings emphasize the
importance of studying spatial inhomogeneities at solid-solid interfaces.  
We hypothesize that atomic lengthscale hot spots at solid-solid interfaces
between the SEI and lithium metal anode surfaces are also the locations where
passivation starts to break down in LELIB.\cite{jungjohann}    While we focus
on the SEI in LELIB, we draw on concepts from SELIB and from electronic
materials.\cite{shluger,liair,fisher,islam,twin}  The models used herein can in
turn be applied to interfaces in SELIB studies.

\begin{figure}
\centerline{\hbox{ (a) \epsfxsize=2.20in \epsfbox{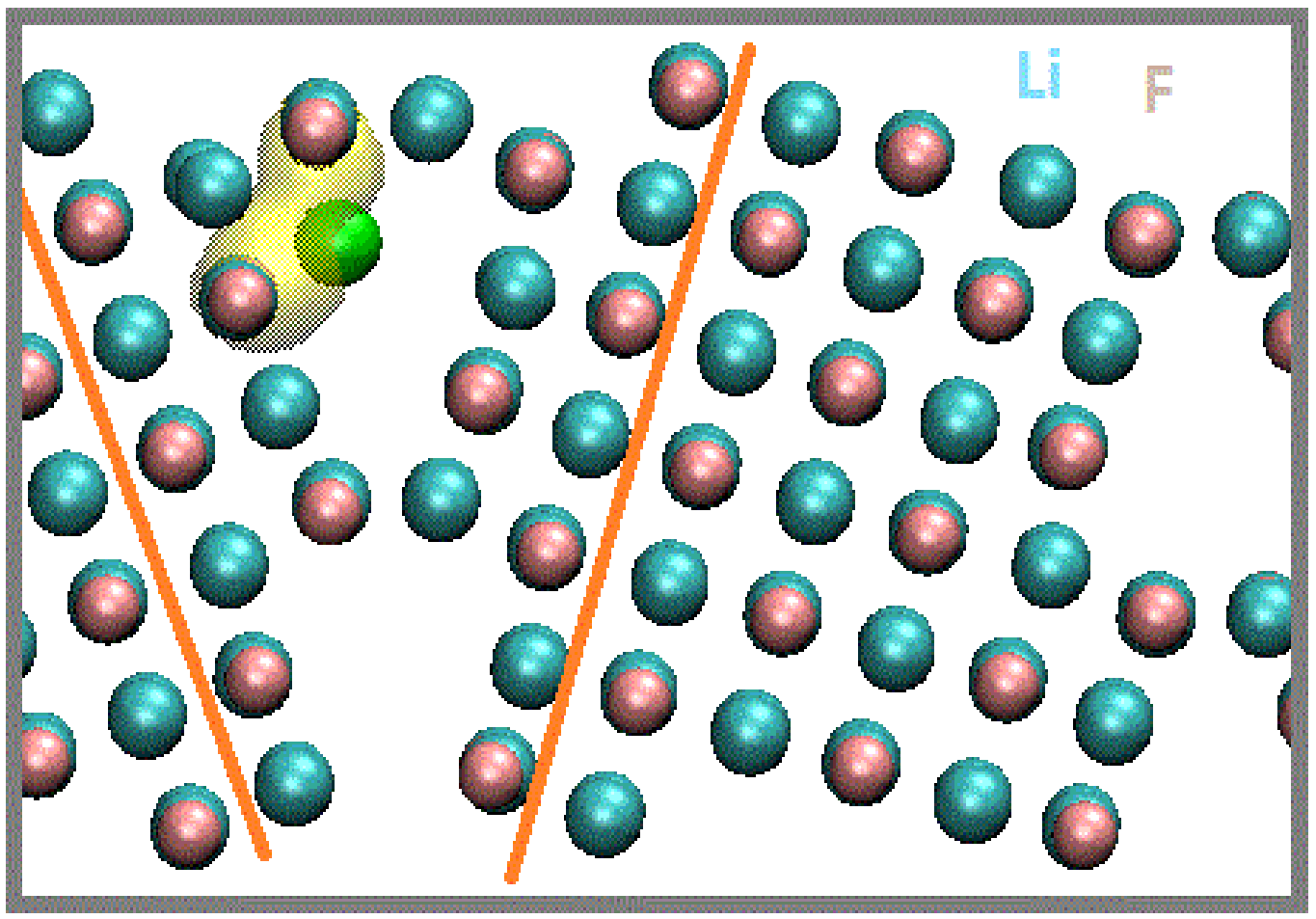} 
		   \epsfxsize=2.20in \epsfbox{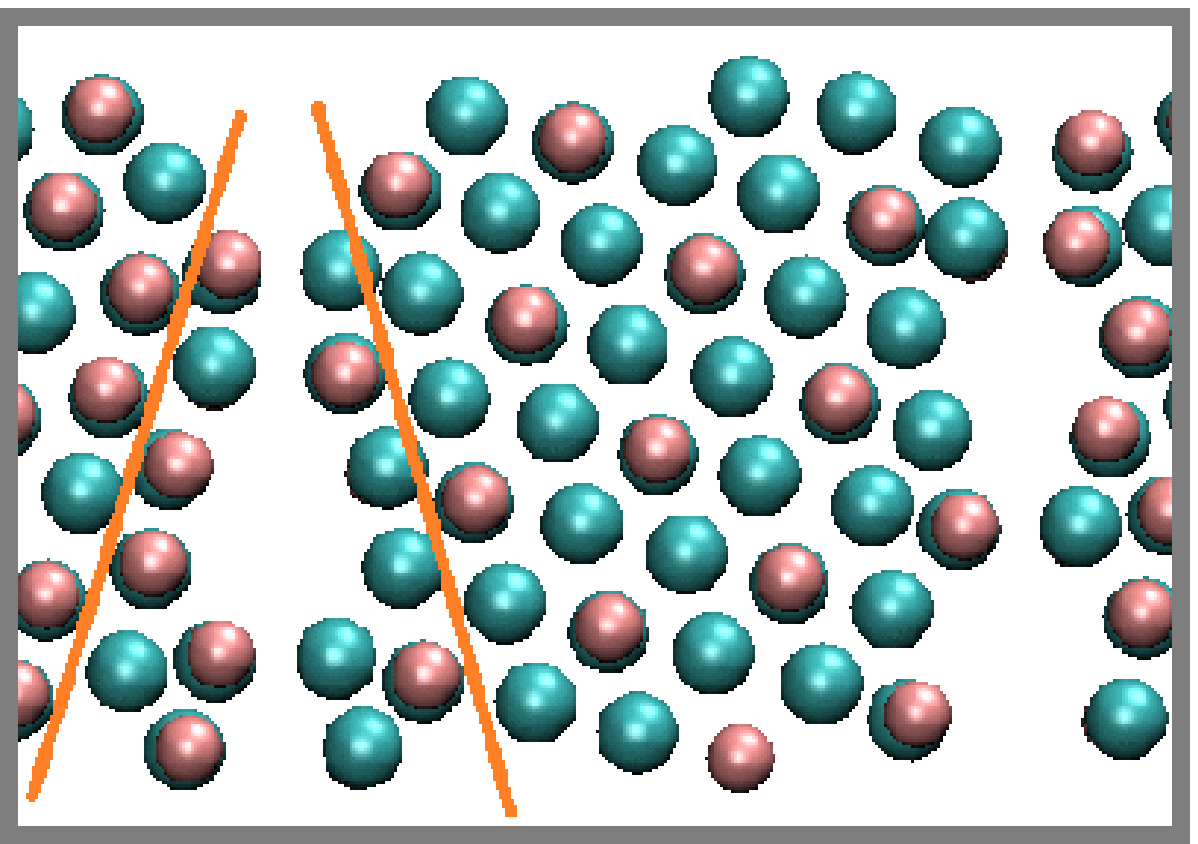} (b)}}
\centerline{\hbox{ (c) \epsfxsize=2.20in \epsfbox{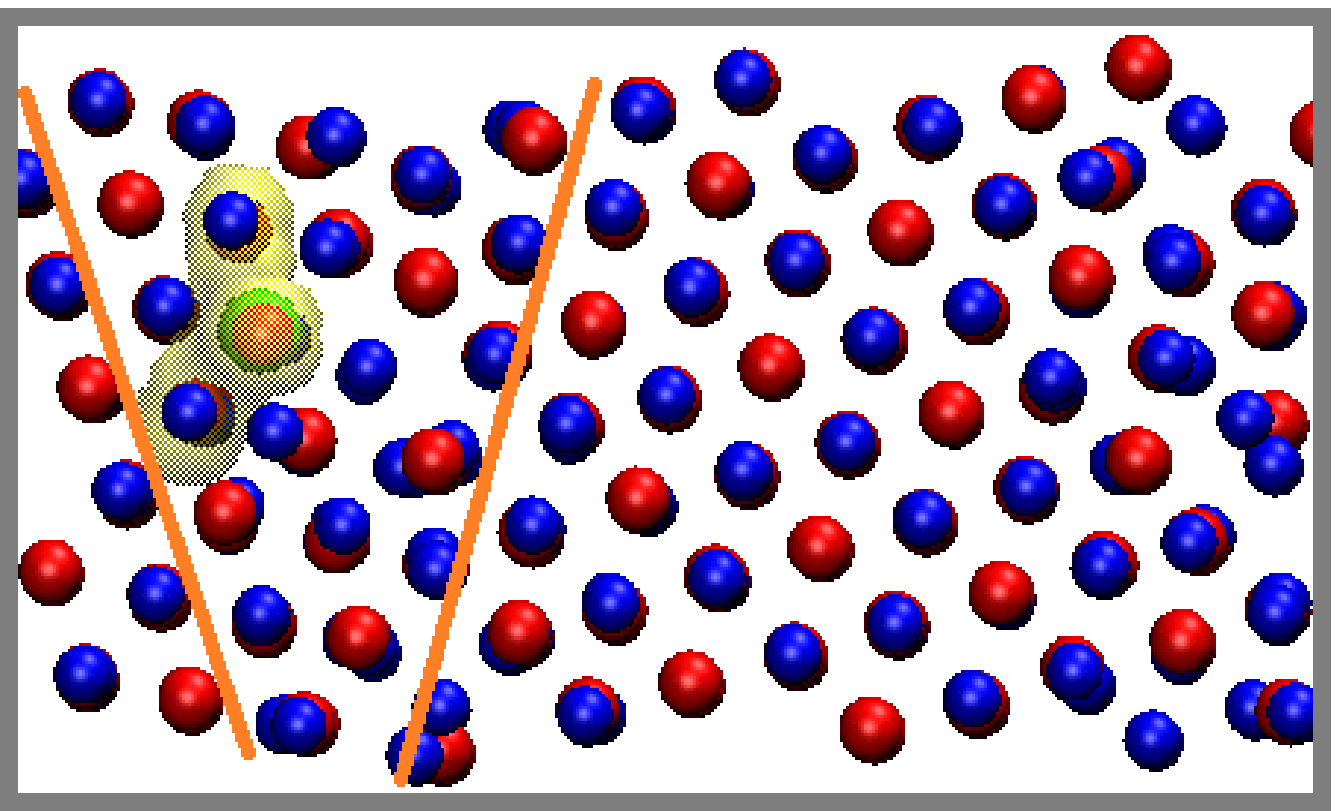} 
		   \epsfxsize=2.20in \epsfbox{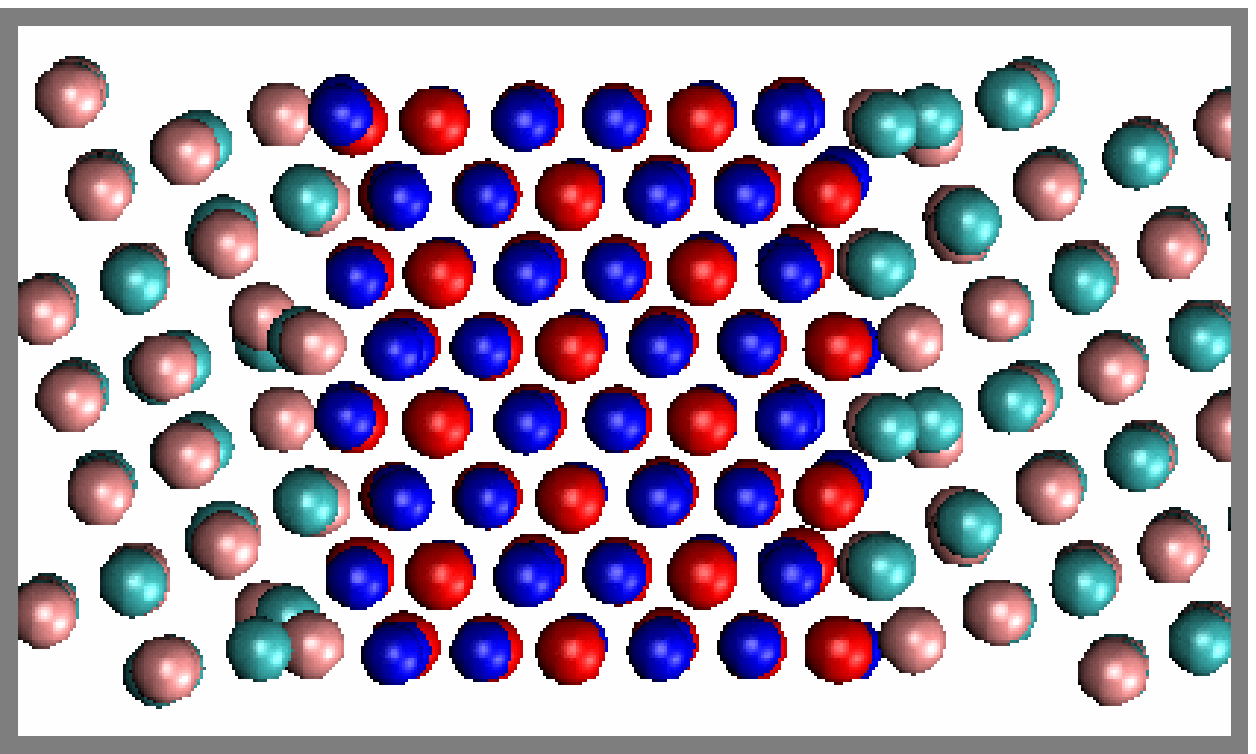} (d)}}
\centerline{\hbox{ (e) \epsfxsize=2.20in \epsfbox{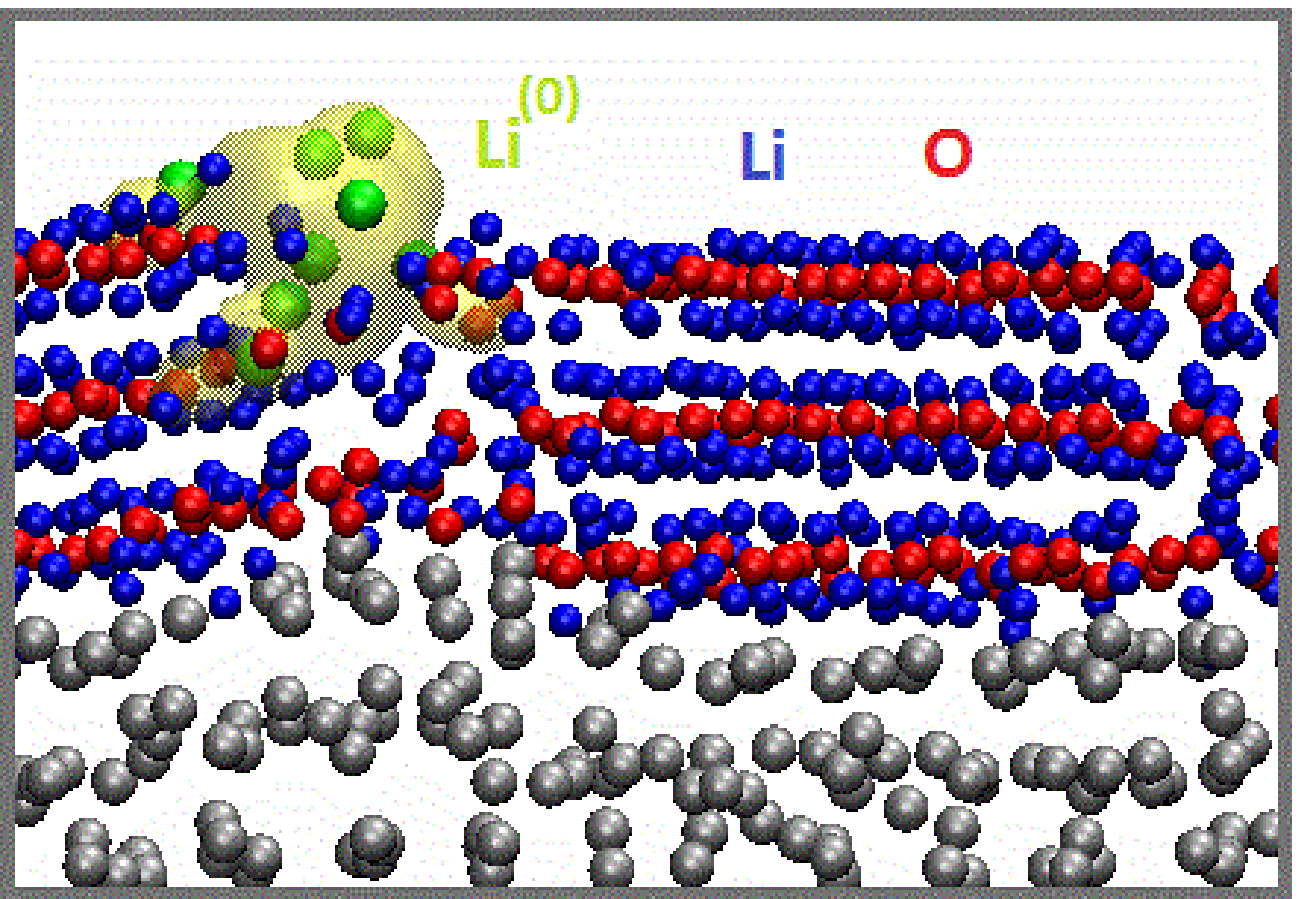} 
		   \epsfxsize=2.20in \epsfbox{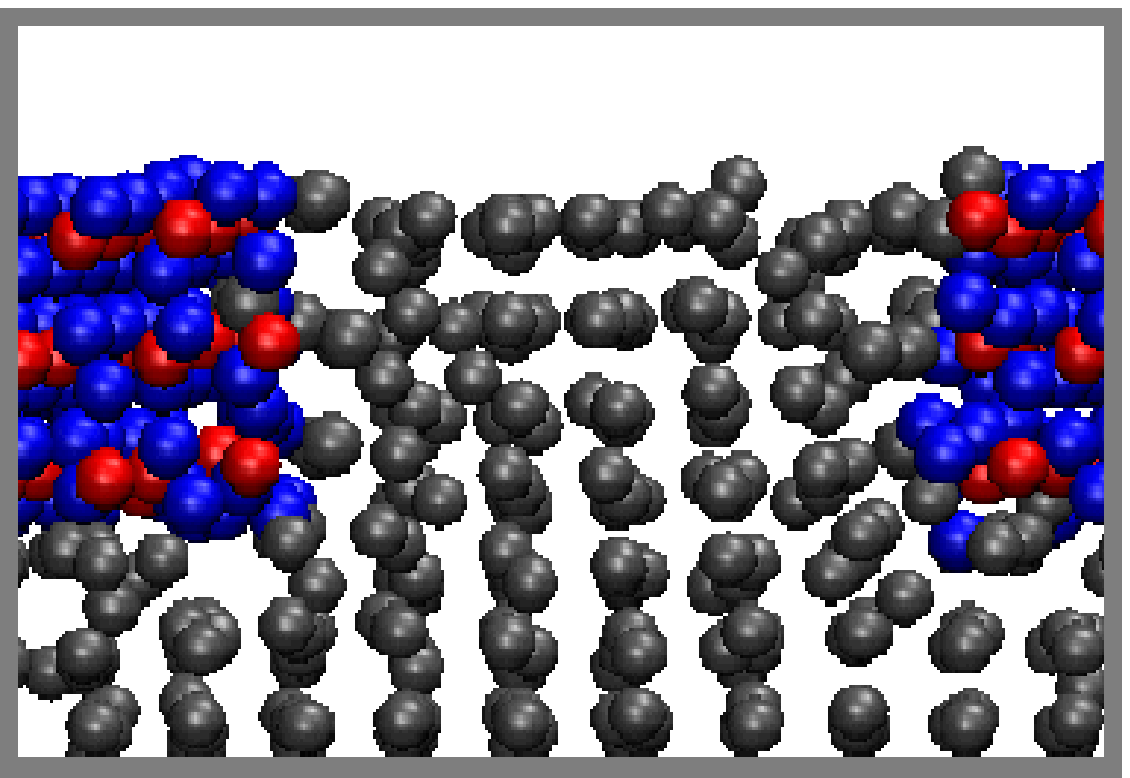} (f)}}
\caption[]
{\label{fig1} \noindent
Representative systems studied in this work.  (a) $\Sigma_5$ grain boundaries
(GB) in LiF; (b) ``16$^o$'' GB in LIF; (c) 16$^o$ GB in Li$_2$O, and (d) GB in
a mixed LiF/Li$_2$O film; (e) Li$_2$O with 16$^o$ grain boundary on Li(s)
surface; (f) Li metal slab inside a $\sim$12~\AA\, crack in Li$_2$O.  
The orange lines indicate the lattice mismatch directions.
Red and pink spheres depict O and F atoms.  The colors of Li atoms depend on
their origins: silver: from Li metal anode; blue: Li$_2$O; cyan: LiF; green:
Li manually added to system and/or Li that has a localized excess electron
according to Bader charge analysis (i.e., Li$^{(0)}$).  The yellow transparent
shapes are contours of excess $e^-$ locations (see text).
}
\end{figure}

While this work is motivated by the science of dendrite formation, it
focuses on lithium metal-induced passivation breakdown on smaller lengthscales
that may cause continuous electrolyte decomposition and may be one of the
root causes of dendrites under adverse conditions.  We apply electronic
structure (Density Functional Theory or DFT) calculations to explore
electron-blockage breakdown and ``incipient lithium filament''
formation in the defect regions of the passivation films.  Understanding the
initial stages of passivation film failure and lithium growth will inform 
early diagnosis, and will potentially lead to self-healing mechanisms that
prevent catastrophic anode failure.  DFT can capture
bond-breaking events and reveal detrimental through-SEI $e^-$ conduction
pathways at sub-nanometer lengthscale; it is complementary to TEM and
phase-field studies.\cite{naturetem,garcia}  
We focus on two SEI components, Li$_2$O and LiF.
Unlike Li$_2$CO$_3$ and other organic SEI components in LELIB,\cite{batt}
Li$_2$O cannot be readily electrochemically reduced by Li(s).  A thin layer
of Li$_2$O is predicted to exist on Li surfaces as the innermost inorganic
SEI layer,\cite{batt} unless LiF, likewise stable, has been deposited first.
This innermost inorganic layer is arguably the most important SEI
component for blocking $e^-$ transport into the liquid electrolyte.\cite{yue}

Our goal is two-fold: to demonstrate that grain boundaries can aid electron
leakage through passivating films, and that cracks initated there can lead to
(sub)-nanoscale Li metal (i.e., incipient filament) growth.  Regarding $e^-$
leakage, we show that Li$^{(0)}$ atoms can reside and diffuse in Li$_2$O
grain boundaries with sufficiently large pore sizes, at $<0.25$~V computed
overpotential (see definition below) vs.~Li$^+$/Li(s) reference.  Li$^{(0)}$
has also been proposed to be $e^-$ carriers in Li$_2$CO$_3$
crystals.\cite{yueli2co3}   This is one possible mechanism responsible for
electron transfer through the
SEI.\cite{tang}  The specific models examined include grain boundaries in
crystalline LiF, Li$_2$O, the heterogeneous boundaries between them
(Fig.~\ref{fig1}a-d),
and thin films containing these grain
boundaries deposited on Li metal anode surfaces with well defined
electronic voltages\cite{solid} (Fig.~\ref{fig1}e).  Regarding Li metal
growth, back-of-the-envelope calculations suggest that Li metal nucleation
can already occur inside a 12~\AA\, pre-existing crack within a Li$_2$O film
at modest computed overpotentials (see the supporting information document,
S.I., Sec.~S1).  Fig.~\ref{fig1}f illustrates a thin Li metal growth inside
a nanometer-wide linear crack.  Other works have focused on the beneficial
effects of heterogeneous interfaces in the SEI.\cite{qi16} 

We stress that grain boundaries in crystals are non-equilibrium structures
and reflect kinetic constraints associated with crystal-growth conditions.
SEI film formation, which occurs in liquid at room temperature, is already
severely kinetically constrained, with many components being thermodynamically
metastable.\cite{batt}  The defects in SEI components are even harder to
characterize experimentally than undefected SEI regions, and models of such
are difficult construct in a systematic way to take proper account of the
kinetic formation constraints.  Cracks that develop in materials are also
clearly kinetically drive.\cite{crack} Nevertheless, it is critical to study
such defects, largely neglected in the literature.  In this work, we adopt
plausible grain boundary models from crystals in the literature\cite{shluger}
and use our own construction (Fig.~S1 in the S.I.).  We show that annealing
these models at high temperature, as has been done in some modeling
publications,\cite{dawson} actually yield ambiguous results.  High temperature
growth are crystal-growth, not SEI-formation, conditions.  To some extent, our
Li$_2$O grain boundary models can be taken as amorphous regions, which we
have postulated to result from electrochemical reduction of other SEI
products.\cite{batt} 

\section*{METHOD}

\subsection*{Construction of the Models}

Our static DFT calculations apply periodically replicated simulation cells,
the Vienna Atomic Simulation Package (VASP) version
5.3,\cite{vasp1,vasp1a,vasp2,vasp3} and the Perdew-Burke-Ernzerhof (PBE)
functional.\cite{pbe}  All simulation cells considered are overall
charge-neutral.  A 400~eV planewave energy cutoff is imposed, except
that a 500~eV cutoff is used when optimizing simulation cell sizes.
Representative simulation cell dimensions, stoichiometries, and Brillouin
zone sampling settings are listed in Table~\ref{table1}.  Other calculations
involve variations on these cells.  Many calculations
involve slab-like simulation cells with a 10-12~\AA\, vacuum region.  In
these cases, the dipole moment correction is applied.\cite{dipole}
In calculations of Li monolayer binding energies, spin-polarized DFT is
applied if there is an odd number of Li atoms in the simulation cell.
Some of these calculations apply the generally more accurate
DFT/HSE06 and DFT/PBE0 functionals.\cite{hse06a,hse06b,hse06c,pbe0}

The two main grain boundaries of interest are the (310)/[100] (henceforth
called ``$\Sigma_5$'') in LiF (001) films, and the one formed by
counter-rotating two Li$_2$O (111) slabs by 16.1$^o$ (simply referred to as
``16$^o$'').  They are chosen because of the stability of LiF (001) and Li$_2$O
(111) surfaces.  Mixed LiF/Li$_2$O boundaries are created by joining the two
surfaces of a LiF (310) slab on to Li$_2$O ($\bar{1}$10) facets
(Fig.~\ref{fig1}d).  In all cases, the $x$-direction is perpendicular to
the grain boundaries.  

The simulation cell containing two $\Sigma_5$ LiF grain boundaries
(Fig.~\ref{fig1}a) is created as follows.  First a LiF crystal at optimal
lattice constants is rotated 18.4$^o$ and cleaved to expose (310) surfaces in
the $z$-direction.  A second, mirror-image  slab is created by reflecting the
first about the $x$-$y$ plane.  The two are pasted together in the periodically
replicated simulation cell to create the two boundaries.   The $x$-dimension
of the cell is varied while using a higher (500~eV) energy cutoff to obtain
the optimal cell length.  There are multiple ways to align these simple cubic
lattice LiF slabs.  The ``coincident site lattice'' (CSL) approach,\cite{fisher}
which posits that the mirror or junction plane is a (310) plane of atoms
common to both slabs, is found to be less energetically favorable than placing
the boundary half a lattice constant from both surfaces, with the slabs shifted
so that a Li always coordinates to a~F (Fig.~\ref{fig1}a).  This configuration
is in fact adopted from Ref.~\onlinecite{shluger}. 

There are limited electronic structure studies of grain
boundaries in fluorite lattice structures of AB$_2$ stoichiometry relevant
to Li$_2$O.\cite{fisher,islam,shluger,brutzel} A simple $\Sigma_5$ grain
boundary is created for Li$_2$O by joining (310) facets.  Since the Li$_2$O
lattice structure is different from that of LiF, the model used in
Fig.~\ref{fig1}a is inapplicable, and the CSL approach to is
applied instead (Fig.~S4 in the S.I.).

For Li$_2$O, the problem with this $\Sigma_5$ grain boundary is that one of
its orthogonal surfaces is (001).  This is the most stable surface for LiF, but
is a high energy surface for Li$_2$O.  The lowest energy facet of Li$_2$O
is (111).\cite{li2o} $\Sigma_5$ is not compatible with a (111) film coating
the Li metal surface.  Instead, taking the (111) direction as the $z$-axis,
we rotate two Li$_2$O slabs in the $x$-$y$ plane by 16.1$^o$ in opposite
directions, join them together in a way to maximize Li-O contacts, and optimize
the $x$ lattice constant as described in the previous paragraph.  This angle
is chosen to give a modest system size with best lattice matching with the
metal surface supercell.  See Fig.~\ref{fig1}c and Table~\ref{table1} for more
details.  In this ``16$^o$'' model, manual insertion of two (but not more)
3-atom Li$_2$O formula units into the grain boundary regions is enegetically
favorable.  As will be discussed, this grain boundary contains sufficient void
space, even after insertion of the two Li$_2$O units, to effectively
accommodate Li$^{(0)}$.  Applying simulated annealing for 12~ps at 500~K and
reoptimizing the structure change the total energy of the simulation
cell by only 0.2~eV.  

\begin{table}\centering
\begin{tabular}{c|r|r|l|r} \hline
system & dimensions & stoichiometry & $k$-sampling & Figure \\ \hline
LiF $\Sigma_5$ GB &  28.53$\times$8.14$\times$12.88 & Li$_{160}$F$_{160}$ &
          2$\times$1$\times$2  & Fig.~\ref{fig1}a \\
LiF 16$^o$ GB &  23.25$\times$10.38$\times$7.05 & Li$_{96}$F$_{96}$ &
          1$\times$2$\times$2  & Fig.~\ref{fig1}b \\
Li$_2$O $\Sigma_5$ GB & 10.37$\times$4.64$\times$22.92 & Li$_{88}$O$_{44}$ &
          2$\times$4$\times$1  & Fig.~S4, S.I. \\
Li$_2$O 16$^o$ GB$^{**}$& 29.50$\times$11.82$\times$8.06 & Li$_{216}$O$_{108}$
        & 1$\times$2$\times$2  & Fig.~\ref{fig1}c \\
LiF/Li$_2$O GB & 27.03$\times$13.12$\times$8.06 & Li$_{180}$O$_{48}$F$_{84}$ &
          1$\times$1$\times$2  & Fig.~\ref{fig1}d \\
LiF GB on Li(s) &  28.53$\times$30.00$\times$19.32 & Li$_{528}$F$_{240}$ &
          1$\times$2$\times$2  & Fig.~\ref{fig3}a-b \\
Li$_2$O GB on Li(s) & 29.50$\times$23.65$\times$32.00 &
        Li$_{804}$O$_{216}$ & 1$\times$1$\times$1 & Fig.~\ref{fig3}c-d \\
LiF/Li$_2$O GB on Li(s) & 27.03$\times$13.12$\times$36.00 &
        Li$_{506}$O$_{80}$F$_{154}$ & 1$\times$1$\times$1 & Fig.~\ref{fig5}a-b\\
LiF$^*$ & 4.07$\times$4.07$\times$24.00 & Li$_{12}$F$_{12}$ &
          4$\times$4$\times$1  & Fig.~\ref{fig2}a \\
Li$_2$O$^*$ & 24.00$\times$3.28$\times$8.06 & Li$_{24}$O$_{12}$ &
          1$\times$4$\times$2  & Fig.~\ref{fig2}b \\
LiF crack on Li(s) & 27.15$\times$20.36$\times$36.00 & Li$_{463}$F$_{140}$ &
          1$\times$1$\times$1  & Fig.~\ref{fig7}a \\
Li$_2$O crack on Li(s) & 23.65$\times$29.25$\times$32.00& Li$_{566}$O$_{106}$ &
          1$\times$1$\times$1  & Fig.~\ref{fig1}f \\
\hline
\end{tabular}
\caption[]
{\label{table1} \noindent
Computational details of representative simulation cells.  ``GB'' refers to
the existence of two matching grain boundaries in the cell.  If unlabelled,
Li$_2$O GB is of the 16$^o$ variety while LiF GB is $\Sigma_5$.  The dimensions
are in \AA$^3$.  $^*$For these systems, we have found that doubling the
density of the $k$-point grid in both lateral dimensions changes the
Li monolayer binding energies by less than 0.05~eV per added Li atom.
$^{**}$Doubling the lateral $k$-point grid changes the total energy by
less than 0.001~eV/atom.
}
\end{table}

For comparison purposes, a similar 16$^o$ grain boundary model for the
LiF surface is also created by rotating LiF (111) slabs by 16.1$^o$
(Fig.~\ref{fig1}b, Table~\ref{table1}).

Finally, mixed LiF/Li$_2$O boundaries are created by joining the two surfaces
of a LiF (310) slab on to Li$_2$O ($\bar{1}$10) facets (Fig.~\ref{fig1}d)
The good lattice matching of these two surfaces
allow cations on one material surface to be coordinated to anions on the
other.  The cell size is optimized as before.  DFT-based molecular dynamics
simulations are further conducted at T=500~K for 7~ps, followed by simulated
annealing to T=100~K in a 3.5~ps trajectory.  This procedure lowers the total
energy, but does not lead to passivation of undercoordinated O$^{2-}$ at
the grain boundary.  Preliminary investigation shows that Li$^{(0)}$ readily
bind to these O$^{2-}$.  To improve passivation,
four LiF dimer units are inserted into voids between the two components so
that all O$^{2-}$ at the interfaces are coordinated to LiF.  Geometry
optimization is re-initiated.  Adding LiF units in this way is found to be
energetically favorable after subtracting the relevant LiF chemical potential,
and appears more fruitful in passivating the grain boundary region with
respect to Li$^{(0)}$ leakage than DFT-based simulated annealing.  
Since SEI formation occurs at room temperature and is kinetically controlled,
it cannot be ruled out that undercoordinated O$^{2-}$ actually exists
at these mixed grain boundaries. Our intention is to construct configurations
that are least hospitable to Li$^{(0)}$ insertion.

The systems described above are periodically replicated; the simulation
cells have no vacuum regions.  To determine the effect of Li metal in their
vicinity, we cut out $\sim 10$~\AA\, films of these materials.
Li ``interlayers'' are added to the bottom of the oxide and/or fluoride films,
such that a Li atom 2~\AA\, exists below each O$^{2-}$ and/or F$^-$ anion
(see Sec. S4 below for rationale).  These films are then placed on Li (001)
or Li (011) surfaces, and the resulting slabs are optimized.  Li (001) and
Li (011) terminations are used interchangeably because they are of similar
surface energies.\cite{holzwarth} We choose the Li(s) facet that gives the
best lattice matching with the inorganic thin film in each case.  The lateral
lattice constants of the soft Li metal are strained to match to the oxide
and/or fluoride.  Afterwards, we also apply strain to the entire systems in
the direction perpendicular to the grain boundaries by various amounts to
mimic curvatures that can develop during lithium plating through the SEI film.
If the model contains a Li$_2$O or LiF film on a Li metal surface, the $z$
direction is perpendicular to the metal surface.  Note that the unstrained
Li$_2$O-on-Li(s) system (Table~\ref{table1}) is in fact first strained by
8.4\% in three successive steps and recompressed to its original cell
dimensions.  This procedure is found to lower the energy of the unstrained
system by 0.6~eV.

For simulation cells containing interfaces with metallic Li electrode slabs,
the true instantaneous electronic voltage ($V_e$) can be computed.
At equilibrium, the Li chemical potential should be consistent with the
electronic voltage.\cite{solid}  We have not applied simulated
annealing to simulation cells with lithium metal anode present.  The
experimental lithium melting point is 180.5$^o$C, or 453.5~K.  Li surfaces
melt at even lower temperatures, making simulated annealing impractical.

\subsection*{Properties and Analysis}

When attempting to insert Li$^{(0)}$ atoms into grain boundaries,
we focus on O$^{2-}$ anions which are coordinated to 6 Li$^+$ or less,
manually add a Li near one of these O$^{2-}$, and optimize the configuration.
The reported binding energies represent the lowest energy obtained in several
insertion attempts.  In perfect crystals of Li$_2$O, each O$^{2-}$ is
coordinated to 8~Li$^+$ and each Li$^+$ to 4~O$^{2-}$.  
We have not considered inserting a Li$^+$ into any of the grain boundaries,
which would create charged simulation cells.  While the excess Li$^+$ may
enhance Li$^+$ concentration and conductivity, it does not address the
failure of passivating film with regard to blocking $e^-$ transport.

The per atom binding energy ($E_{\rm b}$) of Li$^{(0)}$ inside
grain boundaries, or of Li metal films inside cracks, is used to define the
computed overpotential ${\cal V}$.  Thus
${\cal V}$=$-[E_{\rm b}-n E_{\rm Li(s)}]/n|e|$, where $n$ is the number of Li
inserted and $|e|$ is the electronic charge.  This definition of overpotential
broadly corresponds to that of Ref.~\onlinecite{norskov}.  Experimentally,
overpotentials arise from kinetic constraints which are not specified in
this work.  Our definition of ${\cal V}$ merely reflects a convenient way to
describe the insertion energy.

Bader charge analysis\cite{bader} is used to identify localized Li$^{(0)}$
in the insulating Li$_2$O and LiF regions.  Like all charge decomposition
schemes, such analysis is approximate.  It is augmented by examining
the spatial distribution of the excess electron
$\Delta \rho_e({\bf r})$.  Here we first compute the total charge density of a
configuration.  Then the total charge density of the same configuration with
one or more Bader-identified Li$^{(0)}$ removed is computed, keeping the
simulation cell charge-neutral and all other atoms frozen.  Finally, the second
charge density is subtracted from the first.  In the figures, the yellow
transparent shapes represent $\Delta \rho_e({\bf r})$ with density values of
$\sim$0.06~$|e|$/\AA$^3$ or more.  We also plot $\Delta \rho_e(z)$, derived
from integrating $\Delta \rho_e({\bf r})$ over the $x$ and $y$ dimensions.

\section*{RESULTS}
\subsection*{Bulk-like systems with grain boundaries}

First we insert a charge-neutral Li$^{(0)}$ atom into one of the two grain
boundaries in simulation cells mimicking bulk-like SEI materials.
Fig.~\ref{fig1}a depicts an optimized $\Sigma_5$ grain boundary in LiF with
a Li$^{(0)}$ atom coordinated to two F$^-$ anions.  The binding energy is
$-1.41$~eV relative to Li metal cohesive energy (unfavorable).  In other words,
inserting this Li$^{(0)}$ requires a computed overpotential of 1.41~V relative
to the Li$^+$/Li(s) reference.  Bader charge analysis indicates that the
excess $e^-$ is located on the newly added Li. This is confirmed by plotting
the spatial distribution of the excess electron, $\Delta \rho_e({\bf r})$,
in Fig.~\ref{fig1}a.  We have also added a chain of 4~Li atoms in the grain
boundary the periodically replicated cell, which become in effect an infinite
1-D line of Li.  In this case, the computed overpotential needed is lower,
but remains an unfavorable $0.84$~V. Hence thermally-activated insertion of
Li atom into this LiF grain boundary is expected to be rare.  The 16$^o$
grain boundary for LiF (Fig.~\ref{fig1}b) yields a 1.05~V computed
overpotential for inserting one Li$^{(0)}$.  This LiF film would expose a
high energy (111) surface if it were placed on lithium metal surfaces.  Hence
it is not the focus of our studies.

Li~atom insertion into the 16.1$^o$ Li$_2$O boundary (Fig.~\ref{fig1}c) is more
favorable, only requiring an computed overpotential of 0.22~V.  Bader charge
analysis suggests that the excess $e^-$ is centered not at the added Li,
but around another Li$^+$ on the interior grain-boundary surface which is
coordinated to 3 O$^{2-}$ ions in the oxide.  In Fig.~S1 in the S.I., we show
that the SEI film is not a metallic conductor.  The excess $e^-$ resides
inside the band gap.  However, the Li$^{(0)}$ can move by multi-atom hopping.
The diffusion along the 16$^o$ grain boundary is associated with a modest,
0.79~eV barrier.  (See the S.I., Sec. S2.)  Therefore these excess $e^-$ are
reasonably mobile at room temperatures.  We also examine a $\Sigma_5$-like
grain boundary in a Li$_2$O slab.  The computed overpotential for inserting
a Li$^{(0)}$ into the void space there is $1.36$~eV, similar to that associated
with the $\Sigma_5$ boundary in LiF.  The difference between the two Li$_2$O
grain boundaries appears to have a structural origin.
In the optimized configurations after Li$^{(0)}$
insertion, the extra Li$^{(0)}$ in the 16$^o$ grain boundary simulation
cell is 1.85, 1.86, and 1.90~\AA\, from the 3 nearest O$^{2-}$, while the
distances are 1.88, 1.96, and 2.56~\AA\, in the $\Sigma_5$ simulation cell. 
The nearest distances between the added Li and existing Li$^+$ in the lattice
are 2.17 and 2.14~\AA\, in the two cases.  These distances favor Li insertion
into the 16$^o$ grain boundary, which is henceforth our focus.

\begin{figure}
\centerline{\hbox{  \epsfxsize=1.20in \epsfbox{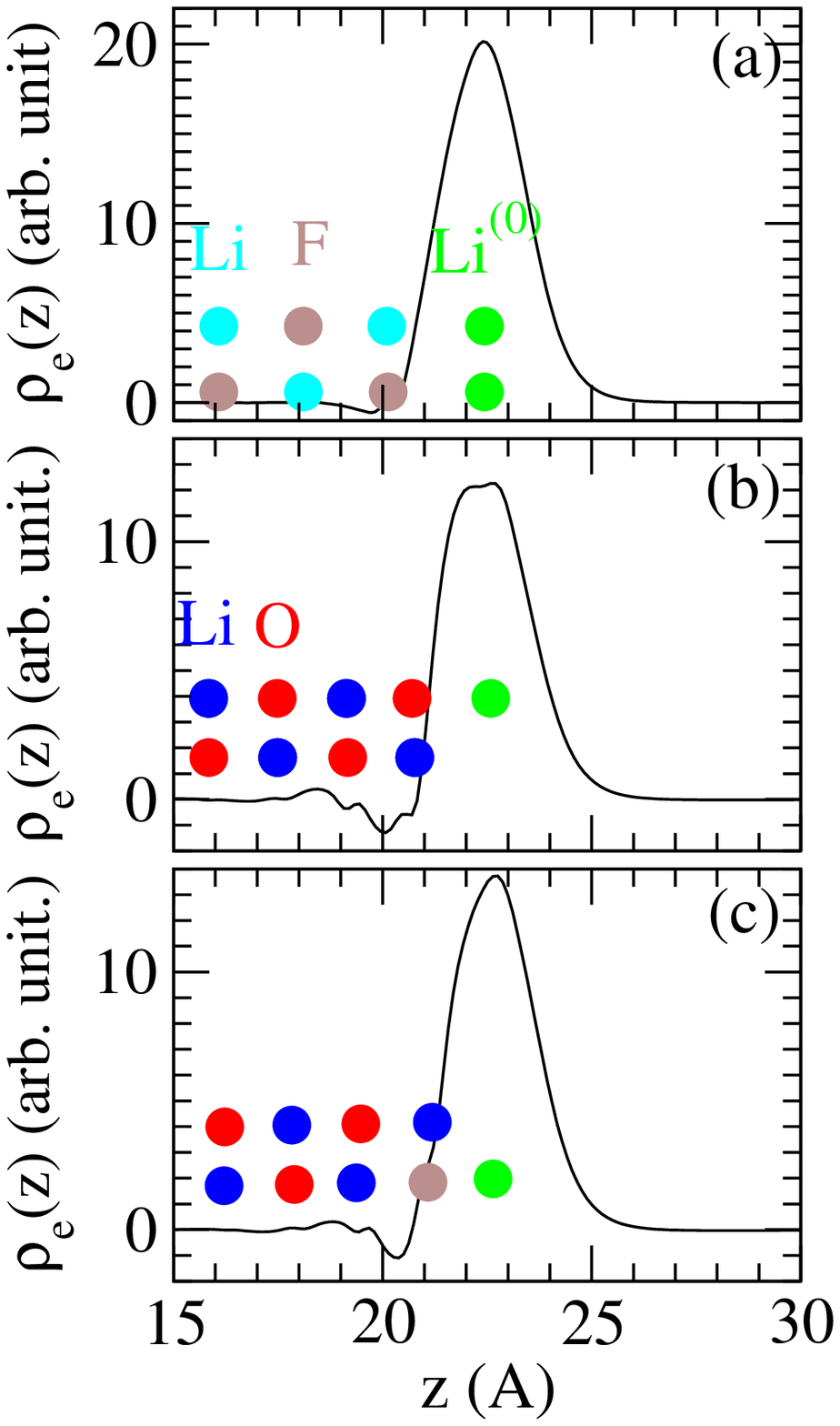}
                   \hspace*{0.05in} \epsfxsize=0.78in \epsfbox{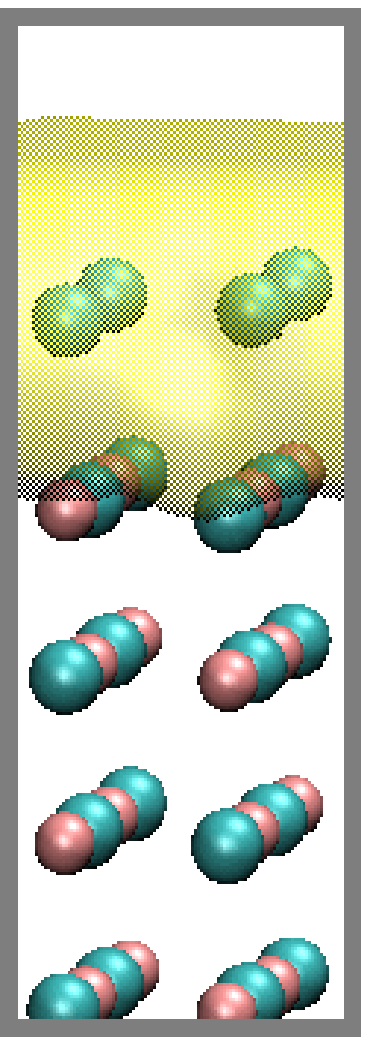}
                    \epsfxsize=0.9in \epsfbox{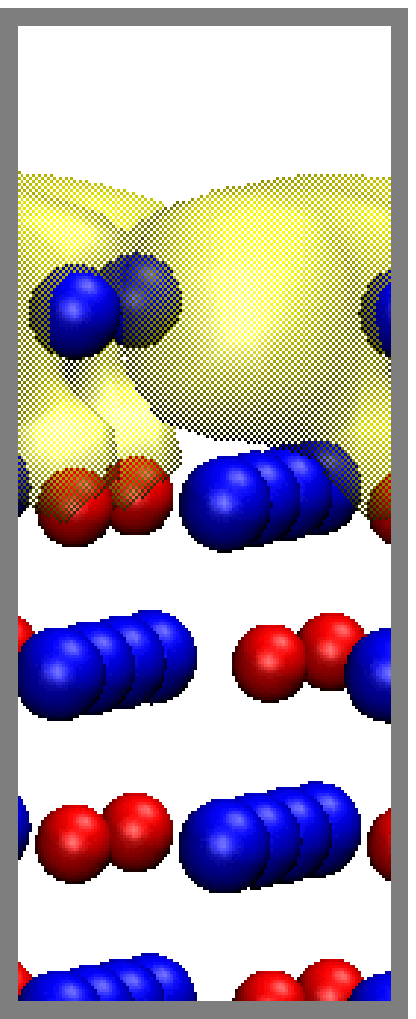}
                    \epsfxsize=0.84in \epsfbox{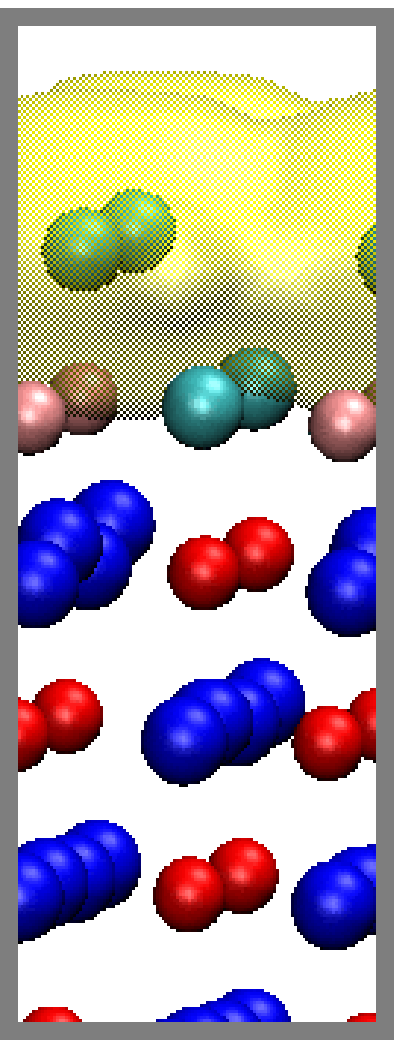} }}
\caption[]
{\label{fig2} \noindent
Excess charge density as a function of the $z$-coordinate, $\Delta \rho_e(z)$,
for Li monolayer adsorption on to (a) LiF (110); (b) Li$_2$O (001);
and (c) Li$_2$O (110) with LiF dimer coating.  The circles indicate the
projection of atoms along the $z$-direction.  Red and pink denote O~and~F;
Li are blue and cyan, and the adsorbed Li monolayer is green.  The three
ball-and-stick images, from second left to right, illustrates systems
(a)-(c) along with contours of excess electron densities.
}
\end{figure}

\subsection*{Surface Energetics}

Given the lack of experimental characterization of defect structures
in the SEI, a qualitative understanding of the above results is needed
to establish they are generally viable, even in disordered regions.
It is possible that defect regions in SEI films may be more accuractely
described as disordered than as defects in crystalline regions.  We propose
that one qualitative difference between LiF and Li$_2$O is that LiF is
a material with a negative electron affinity,\cite{shluger} while Li$_2$O
supports surface states that can accommodate excess $e^-$.  Electron affinity
is relevant because the interior surfaces inside a grain boundary confine a
vacuum region.  We quantify this effect by comparing an adsorbed monolayer
of Li metal on LiF (001) and Li$_2$O ($\bar{1}$10) (Fig.~\ref{fig2},
Table~\ref{table2}).  Here the model Li monolayer, as opposed
to a well-isolated Li~atom on the surface, allows the use of smaller 
simulation cells, and therefore more costly hybrid DFT functionals which more
accurately describe electron localization effects. 

The computed overpotentials for adding the Li monolayer are 0.45~V and 0.23~V,
respectively.  Even Li$_2$O (111), the most stable Li$_2$O facet, exhibits a
smaller computed overpotential towards Li monolayer adsorption than LiF (001)
(Table~\ref{table2}).  Two other Li$_2$O facets actually favor Li monolayer
adsorption (Table~\ref{table2}).  There are only two (111) surfaces in Li$_2$O
crystals, and mulitple facets must be exposed at its grain boundaries.

\begin{table}\centering
\begin{tabular}{l|r|r|r|r} \hline
facet & (111) & (310) & ($\bar{1}$10) & ``16$^o$'' \\ \hline
sur.~energy & 0.54 & 1.11 &  0.94  & 1.06 \\
Li monolayer & $-0.346$ & 0.028 & $-0.228$ & 0.036 \\
\hline
\end{tabular}
\caption[]
{\label{table2} \noindent
Surface energies (J/m)$^2$ and binding energies of Li monolayer relative
to Li bulk chemical potential (eV/Li) for selected Li$_2$O facets.
Positive Li monolayer adsorption energies mean favorable adsorption.  No
attempt is made to remove strain in the Li adatom films.\cite{holzwarth}
}
\end{table}

Fig.~\ref{fig2}a-b compare the integrated differential charge densities
($\Delta \rho_e(z)$) after adding an Li monolayer to LiF (001) and
Li$_2$O ($\bar{1}$10).  On LiF (001), $\Delta \rho_e(z)$ is almost entirely
localized on the Li adatoms.  On Li$_2$O ($\bar{1}$10), a more substantial
part of $\Delta \rho_e(z)$ has leaked on to the top layer O$^{2-}$ ions.
The three-dimensional contour plots in Fig.~\ref{fig2} further confirm this
difference.  Using the HSE06 and PBE0 functionals yield energy differences
that are within 50~meV of those computed using the PBE functional, and
$\Delta \rho_e(z)$ profiles that are indistinguishable from PBE predictions.

We have also coated the Li$_2$O surface with a LiF monolayer
(Fig.~\ref{fig2}c). The computed overpotential associated with adding Li to
this surface, 0.64~V, is even less favorable than that on bare LiF (001).
LiF can originate from decomposition of PF$_6^-$ counter-ions found in organic
solvent-based electrolytes, but is more rapidly released when fluoroethylene
carbonate (FEC) additive molecules are present.\cite{fec1,fec2,fec3}
LiF appears to play a special role in electrode surface passivation.

\begin{figure}
\centerline{\hbox{ (a) \epsfxsize=2.20in \epsfbox{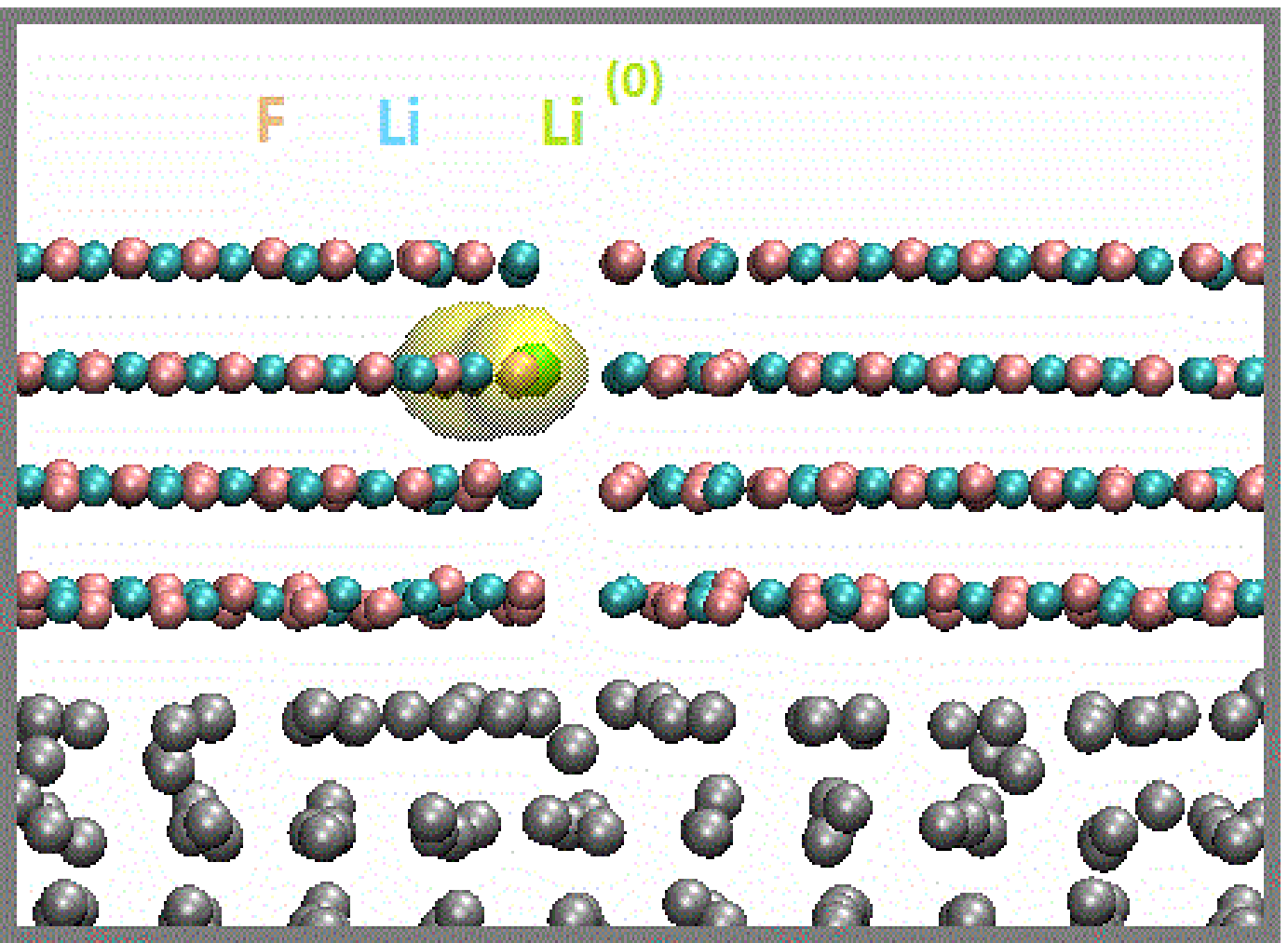} 
		   \epsfxsize=2.20in \epsfbox{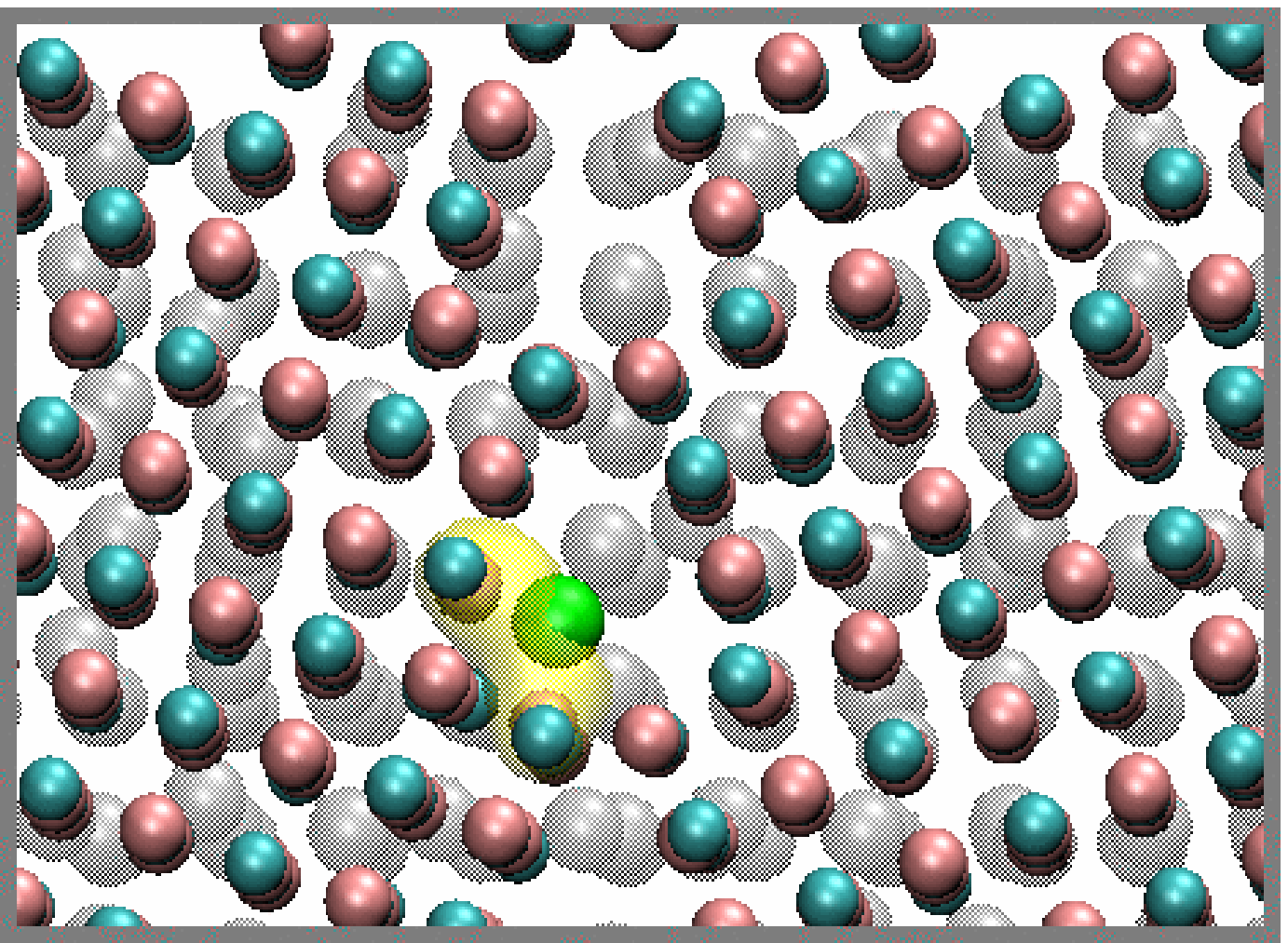} (b)}}
\centerline{\hbox{ (c) \epsfxsize=2.20in \epsfbox{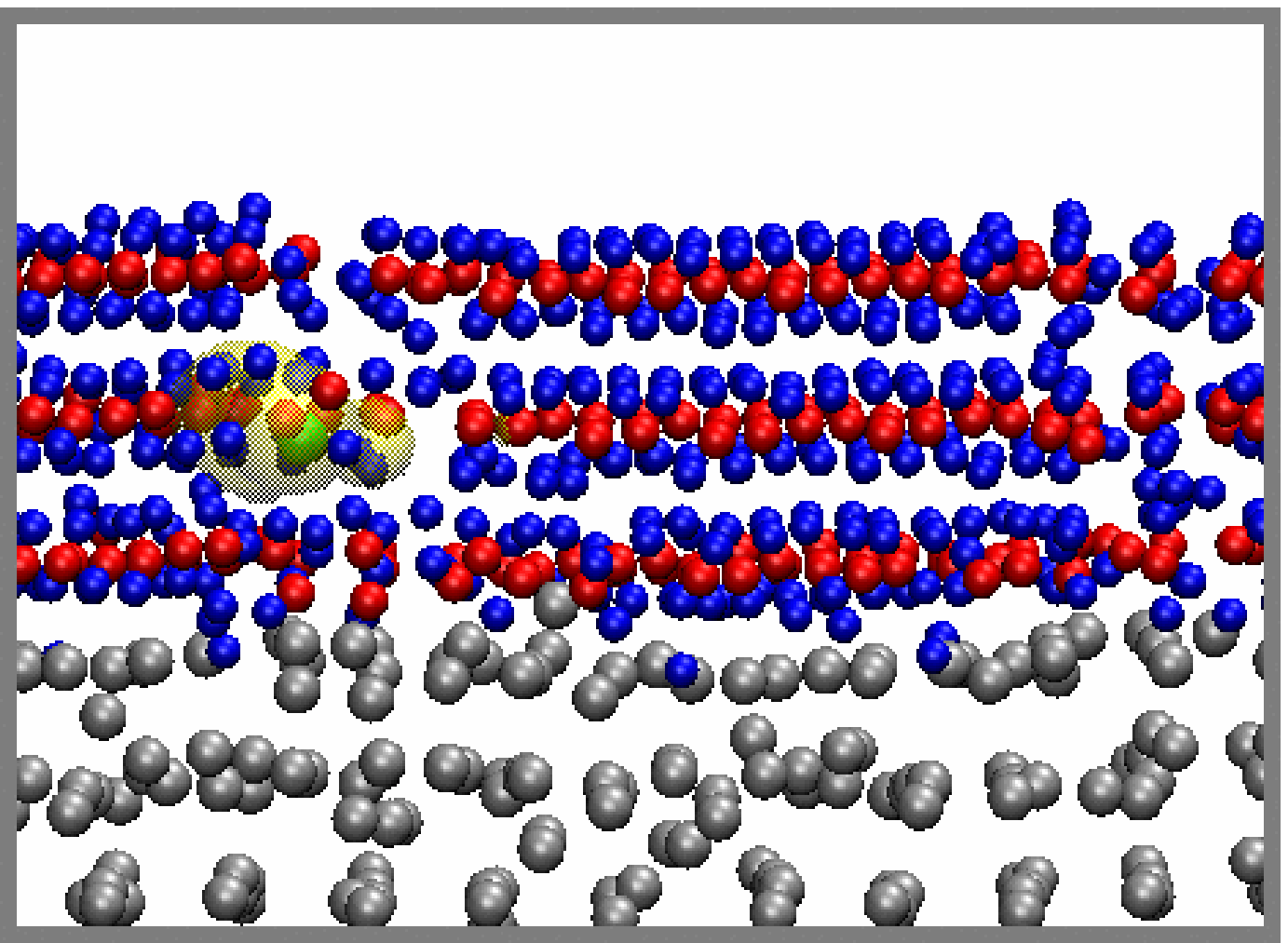} 
		   \epsfxsize=2.20in \epsfbox{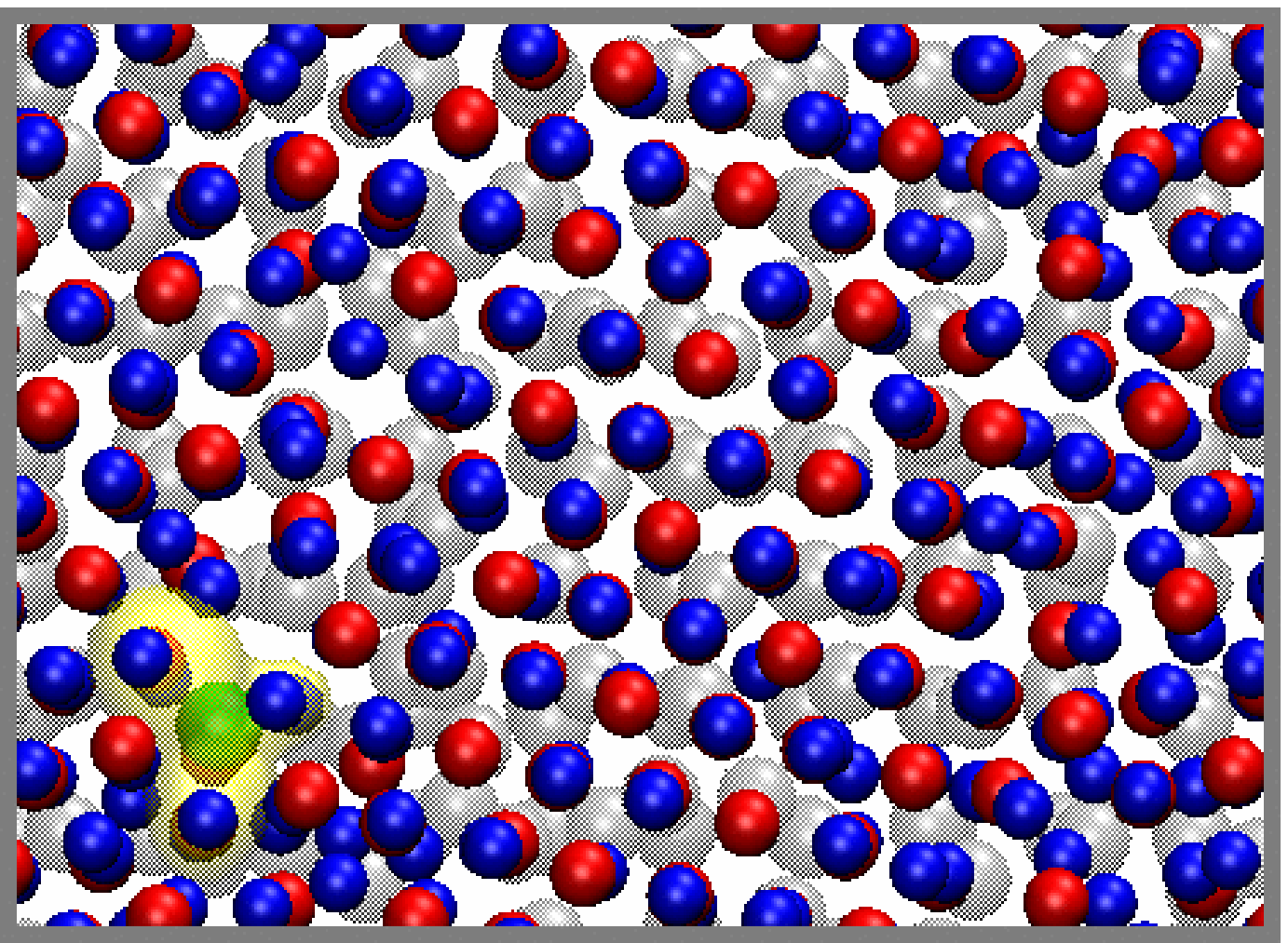} (d)}}
\centerline{\hbox{ (e) \epsfxsize=2.20in \epsfbox{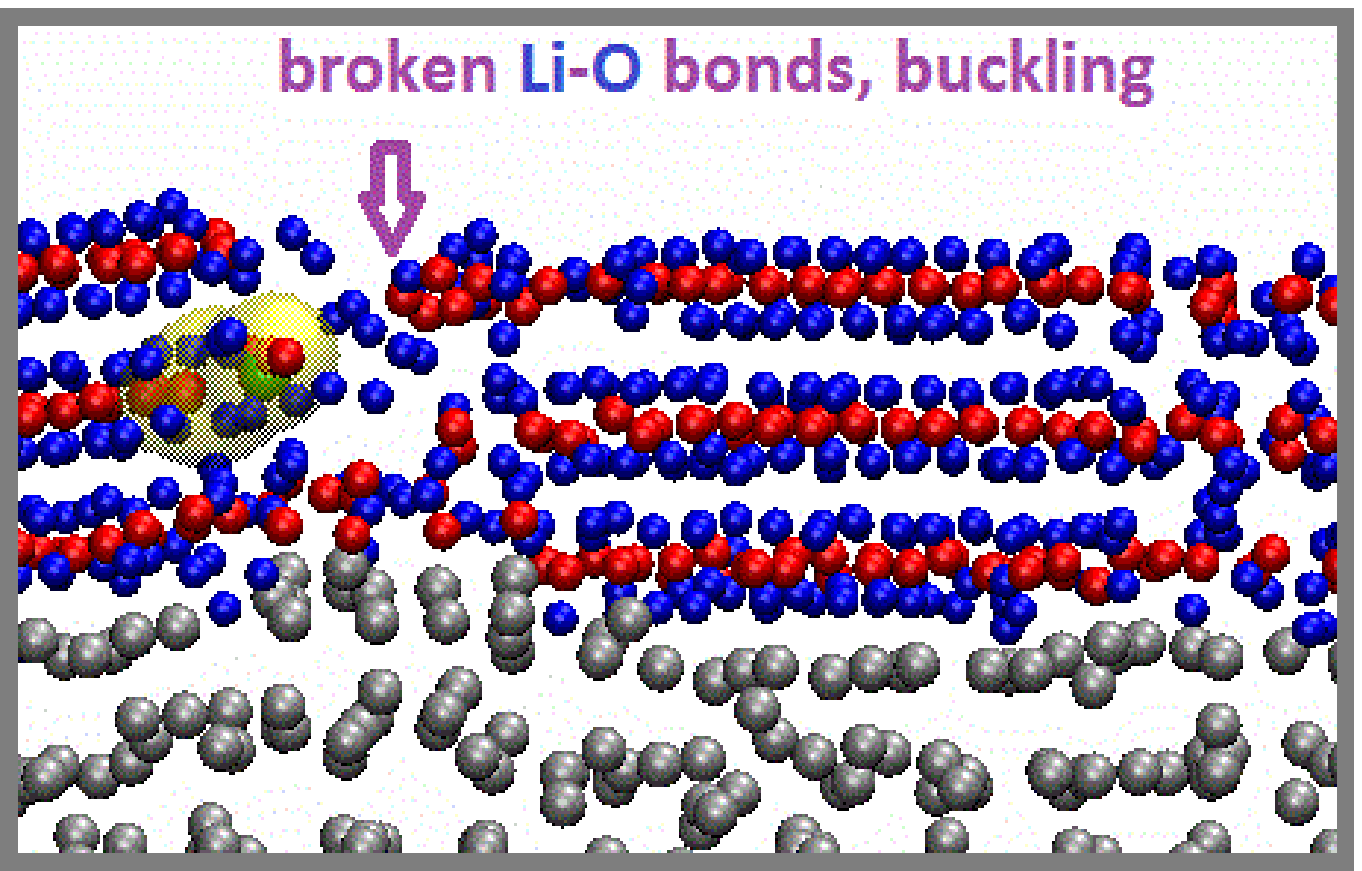} 
		   \epsfxsize=2.20in \epsfbox{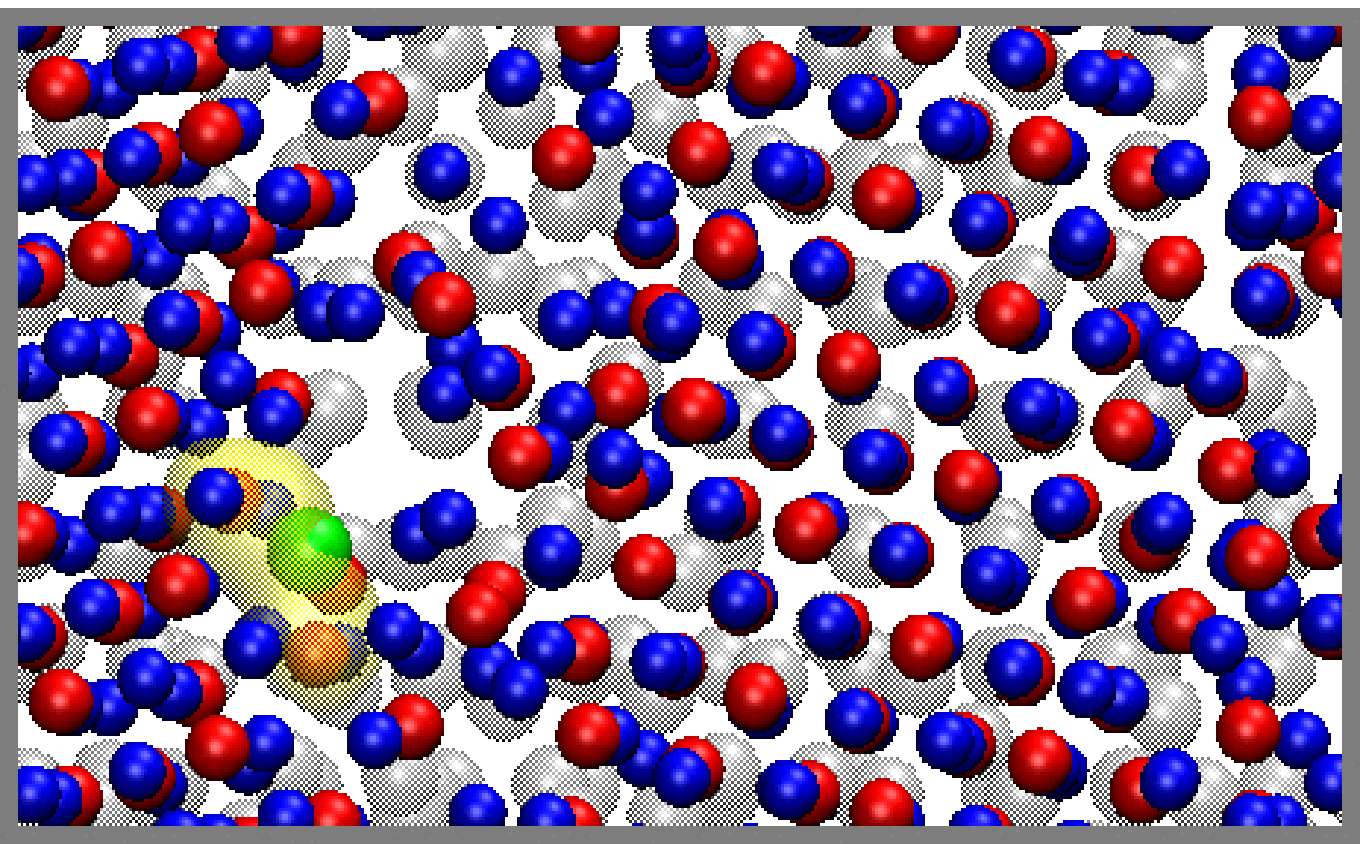} (f)}}
\caption[]
{\label{fig3} \noindent
Side and top views of films with grain boundaries on Li metal surfaces.
(a)-(b): LiF; (c)-(d): Li$_2$O; (e)-(f): Li$_2$O strained by $\sim$12\%.
For color key, see Fig.~\ref{fig1}.  In the top view panels, the
Li metal underneath are depicted as transparent.
}
\end{figure}

\subsection*{Grain boundaries in films on lithium metal}

Whether Li$^{(0)}$ can reside inside grain boundaries, however, ultimately
depends on the interface-modified Fermi level of the anode in contact with
the surface film.  Next we attempt to insert Li$^{(0)}$ in thin films
with grain boundaries deposited on Li metal.  We avoid adding Li near the Li
metal surface, where it will simply be absorbed into the metal electrode, or
on the outer film surface.  Fig.~\ref{fig3}a-b show two perspectives of
a $\sim$10~\AA\, thick LiF film, cut from Fig.~\ref{fig1}a deposited on Li
metal.  A Li$^{(0)}$ is added approximately 6.1~\AA\, from the center of mass
of the top Li layer in the anode (Fig.~\ref{fig3}a).  The computed
overpotential is 1.49~V, similar to the LiF model without Li metal
(Fig.~\ref{fig1}a).  Bader analysis indeed reveals that the added Li has
an excess $e^-$.

The Li$_2$O film on Li(s) (Fig.~\ref{fig3}c-d) already contains two Li$^{(0)}$
atoms in one of its grain boundaries after geometry optimization; manually
inserting Li$^{(0)}$ is not needed.  One Li$^{(0)}$ is on the outer surface
coordinated only to one O$^{2-}$.  Of more interest is the other Li,
coordinated to two O$^{2-}$, halfway through the Li$_2$O film (5.6~\AA\, from
the Li metal surface).  Removing the latter Li$^{(0)}$ reveals that it has an
computed overpotential of only 0.25~V.  This overpotential is strongly
strain-dependent, and falls to a mere 0.1~V upon applying a 1.7\% strain.  The
existence of a Li$^{(0)}$ inside the Li$_2$O film makes this film
non-passivating.

Next, a 3.5~\AA, or about 12~\%, tensile strain is applied to the Li$_2$O
cell with 16$^o$ grain boundaries in the $x$-direction, in 3 successive
increments each followed by geometry optimization (Fig.~\ref{fig3}e-f).  This
mimics possible surface curvatures arising from Li plating during charging.
Silicon anodes are known to expand volumetrically by up to 400\% during
charging, while graphite can expand its c-axis spacing by $\sim$10\%.  There is
less documented data about local strain on SEI-covered Li surfaces.  We choose
a $12\%$ expansion as the outer limit.  Substantial bulging of Li metal anode
surface, and buckling of the film above it, are observed upon applying the
strain (Fig.~\ref{fig3}e).  Multiple Li-O ionic bond cleavage events occur
at one of the two grain boundaries, leaving a sub-nanometer-sized crack.
Inserting 9 Li atoms into the crack is found to require no computed
overpotential, yielding a small Li particle on the outside surface of the
Li$_2$O film (Fig.~\ref{fig1}e).  This suggests that incipient lithium metal
dendrites can nucleate on sub-nanometer defect features inside cracks in 
Li$_2$O.  Note that we do not claim to separate the effects
of strain and broken bonds.  As mentioned above, our grain-boundary
models can be thought of as amorphous regions in the SEI.

\begin{figure}
\centerline{\hbox{ \epsfxsize=2.20in \epsfbox{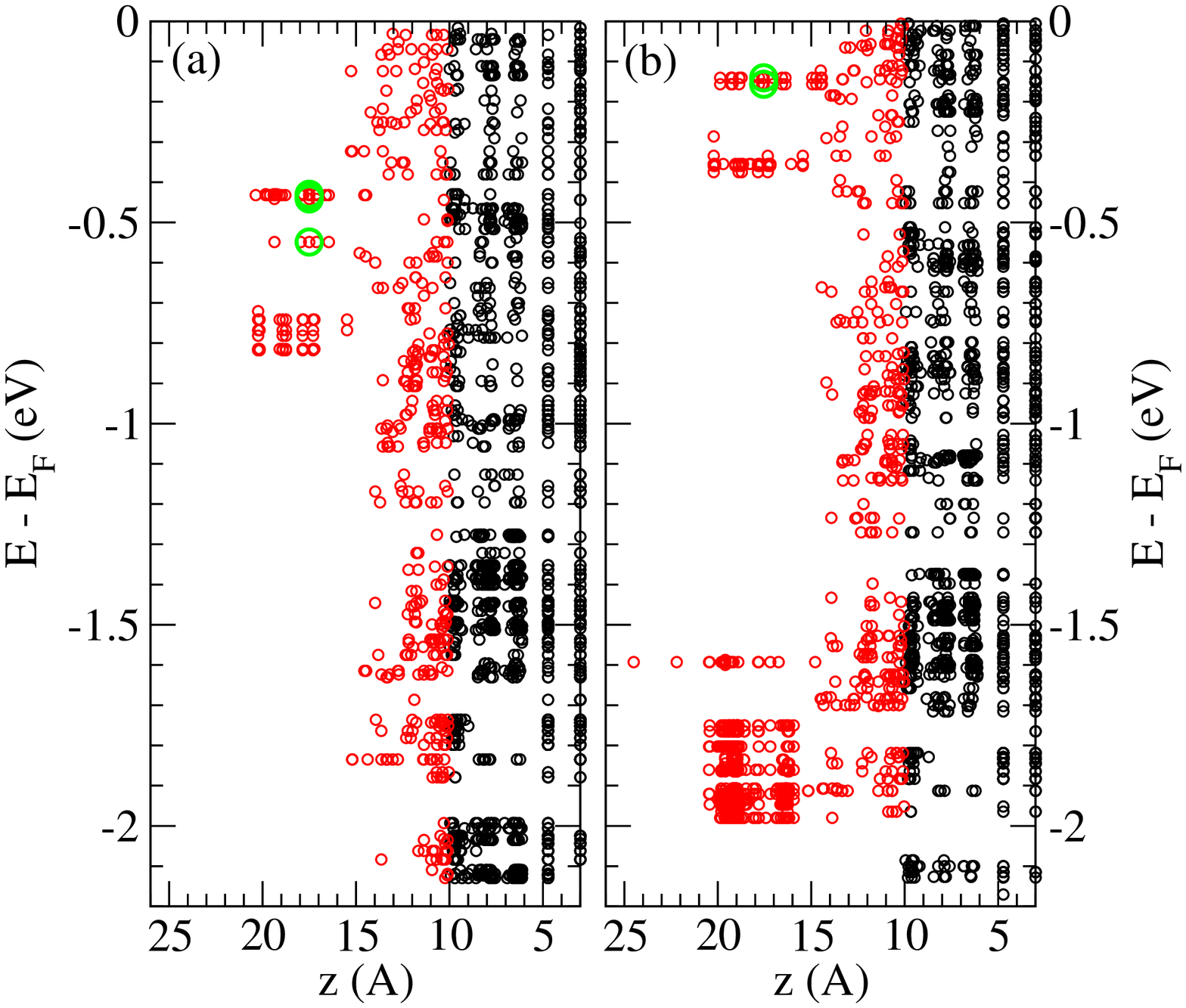} \hspace*{0.1in}
	 \epsfxsize=2.40in \epsfbox{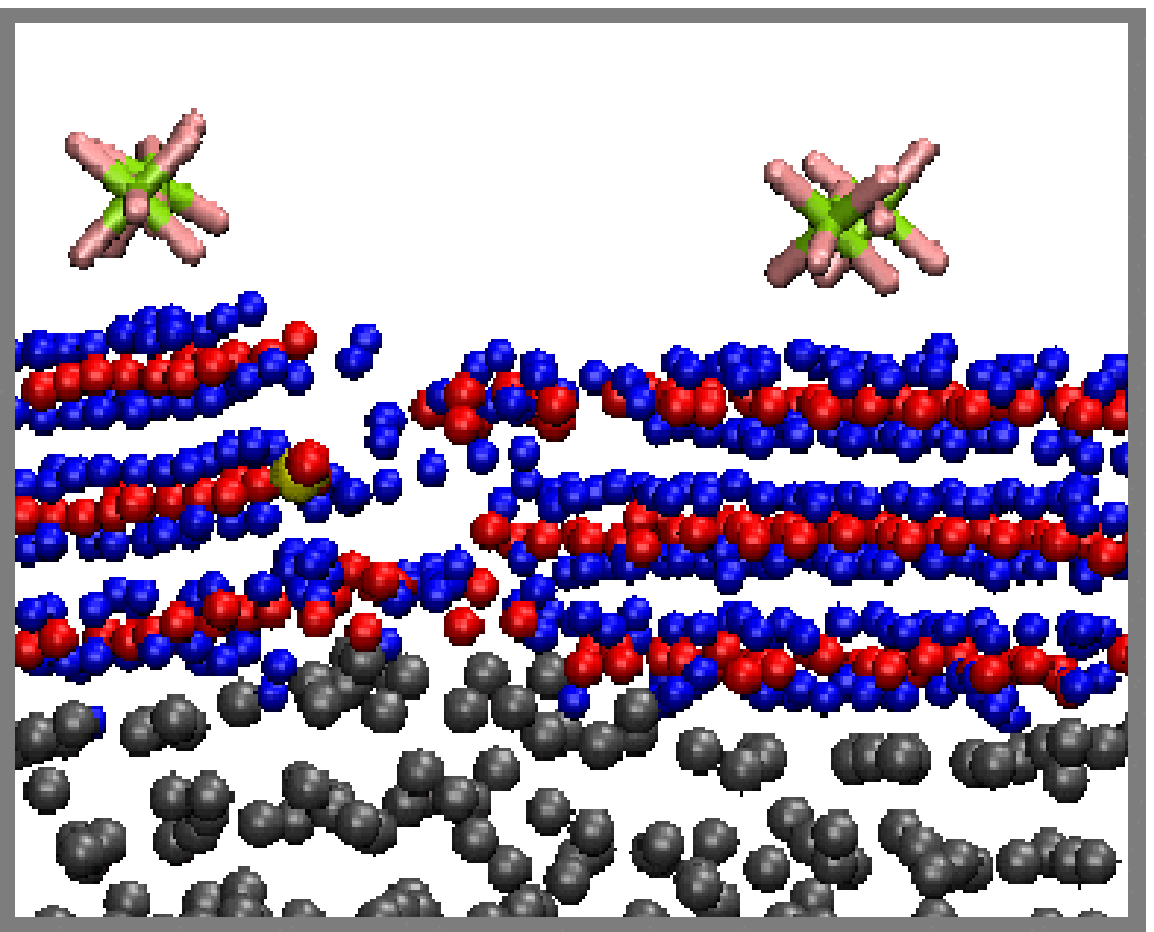} (c) }}
\caption[]
{\label{fig4} \noindent
(a)-(b) Orbital energies along the $x$-direction for the 3.5-\AA\,-stretched
Li$_2$O-covered Li surface, with and without 4 PF$_6^-$ anions
(Fig.~\ref{fig3}e and panel (c)), respectively.  Black and red circles
represent $z$$<$10~\AA\, and $z$$>$10~\AA\, contributions, while green
circles denote orbitals associated with the Li$^{(0)}$ inside the Li$_2$O
film identified using Bader charge analysis.  (c) Similar to Fig.~\ref{fig3}e,
but with 4 PF$_6^-$ added.  P and F are in green and pink.
}
\end{figure}

We also address the instantaneous electronic voltage ($V_e$) of the systems
we have studied.  The work function of the Li metal electrode, modified by
the thin film, is the absolute Fermi level of the electrode referenced to
vacuum.  Dividing the work function by $|e|$ and subtracting 1.37~V yields
$V_e$ referenced to Li$^+$/Li(s).\cite{solid} At equilibrium, $V_e$ should
be equal to the Li=(Li$^+$+$e^-$) chemical potential-derived ``equilibrium
voltage,'' or ''computed overpotential,'' discussed earlier.

The LiF- and Li$_2$O-coated Li metal surfaces, respectively Fig.~\ref{fig3}a-b
and Fig.~\ref{fig3}c-d, exhibit $V_e$=0.07~V and 0.16~V.  These computed
voltages are very close to the Li-plating potential of 0~V vs.~Li$^+$/Li(s).
Fig.~\ref{fig4}a depicts the the energies of Kohn-Sham orbitals along the
$z$-direction in the Li$_2$O film strained by 3.5~\AA\, (Fig.~\ref{fig3}e-f).
Green circles represent orbitals associated with the Li identified by Bader
charge analysis to be a Li$^{(0)}$.  The relevant, occupied localized orbital
on the Li$^{(0)}$ is about 0.4~eV below the lithium metal Fermi level.

In our model, the Li(s)/SEI-film and SEI/vacuum interfaces
separate Li(s) from the vacuum region.  In more realistic electrochemical
systems, the outer surfaces of the inorganic films are in contact with
organic SEI components and/or liquid electrolytes.  To maintain a 0.0~V
applied voltage in those more realistic systems, the charge distribution in
the SEI film, electrolyte, and their interfaces may be slightly different.
Hence we need to show that varying $V_e$ has only small effects on the 
existence of Li$^{(0)}$ inside the SEI.  

Fig.~\ref{fig4}c depicts four PF$_6^-$ adsorbed on the outer oxide surface.
PF$_6^-$ retains all its negative charge in vacuum, without the need for
solvation.  In the charge-neutral simulation cell, PF$_6^-$ induce
compensating positive charges on the Li(s) surface.  The dipole surface
density created exerts a large electric field and raises $V_e$ by
2.66~V.\cite{solid}  But this electric field is found to raise the energy
of the Li$^{(0)}$ orbital by only 0.3~eV (Fig.~\ref{fig4}b), even though
$V_e$ is raised by several times that much.  The Li$^{(0)}$ orbital remains
occupied (Fig.~\ref{fig4}b).  The reason is that the $e^-$ is localized
away from the thin film-vacuum interface, and does not experience
the entire voltage drop through the inorganic film.  Note that if the model
does not contain an interface, but is a bulk simulation cell (vanishing
electric field gradient approximation), DFT would erroneously predict that
the voltage has no effect on a charge-neutral defect like Li$^{(0)}$.

\subsection*{Grain Boundary in Mixed Li$_2$O/LiF Films}

\begin{figure}
\centerline{\hbox{ (a) \epsfxsize=2.20in \epsfbox{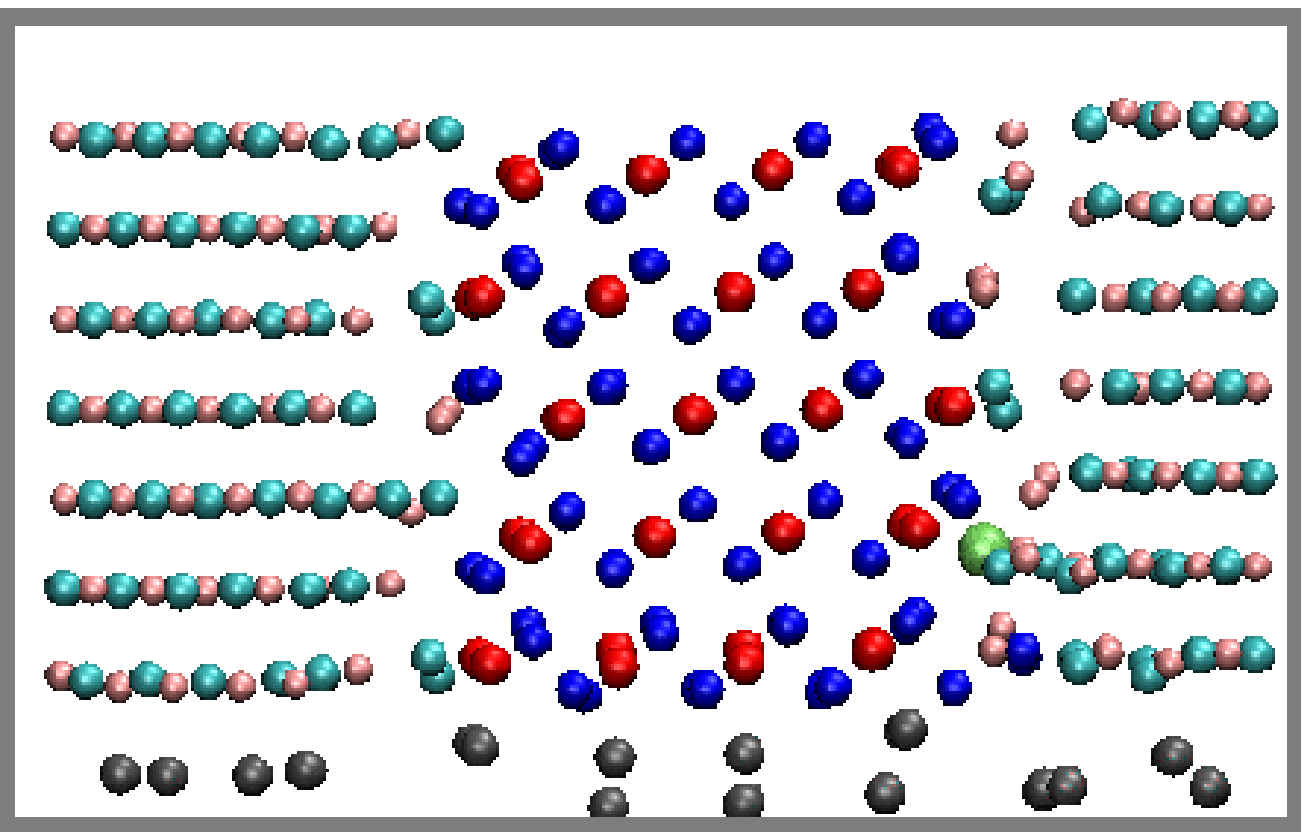}
                   \epsfxsize=2.20in \epsfbox{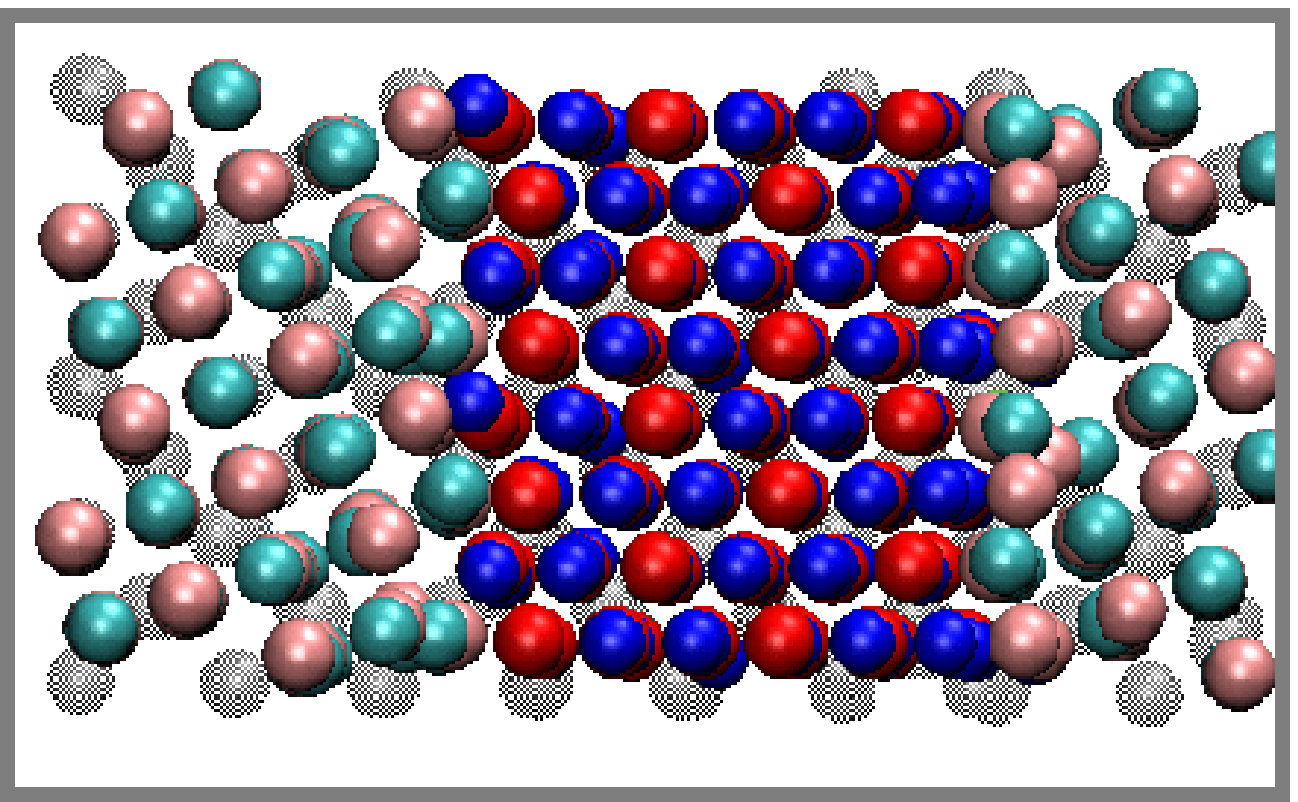} (b)}}
\centerline{\hbox{ (c) \epsfxsize=2.20in \epsfbox{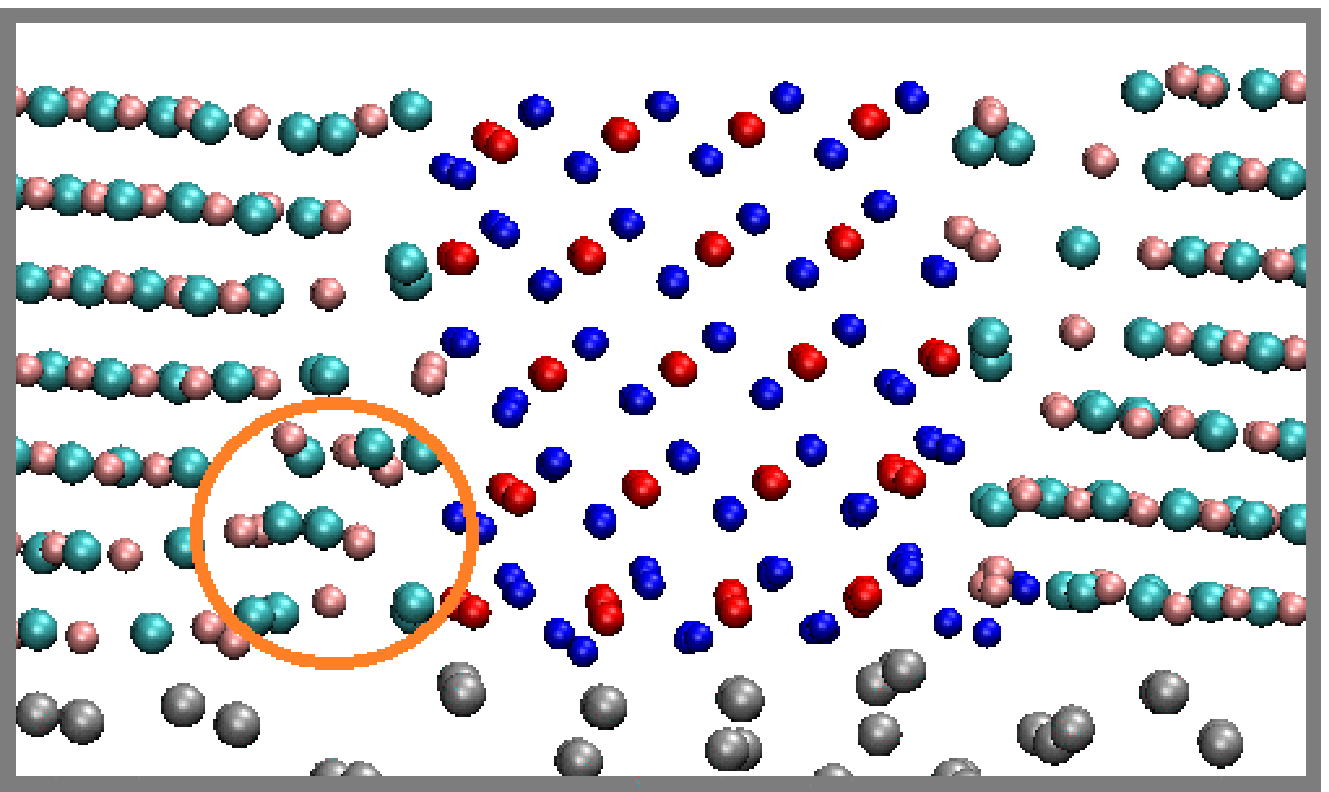}
                   \epsfxsize=2.20in \epsfbox{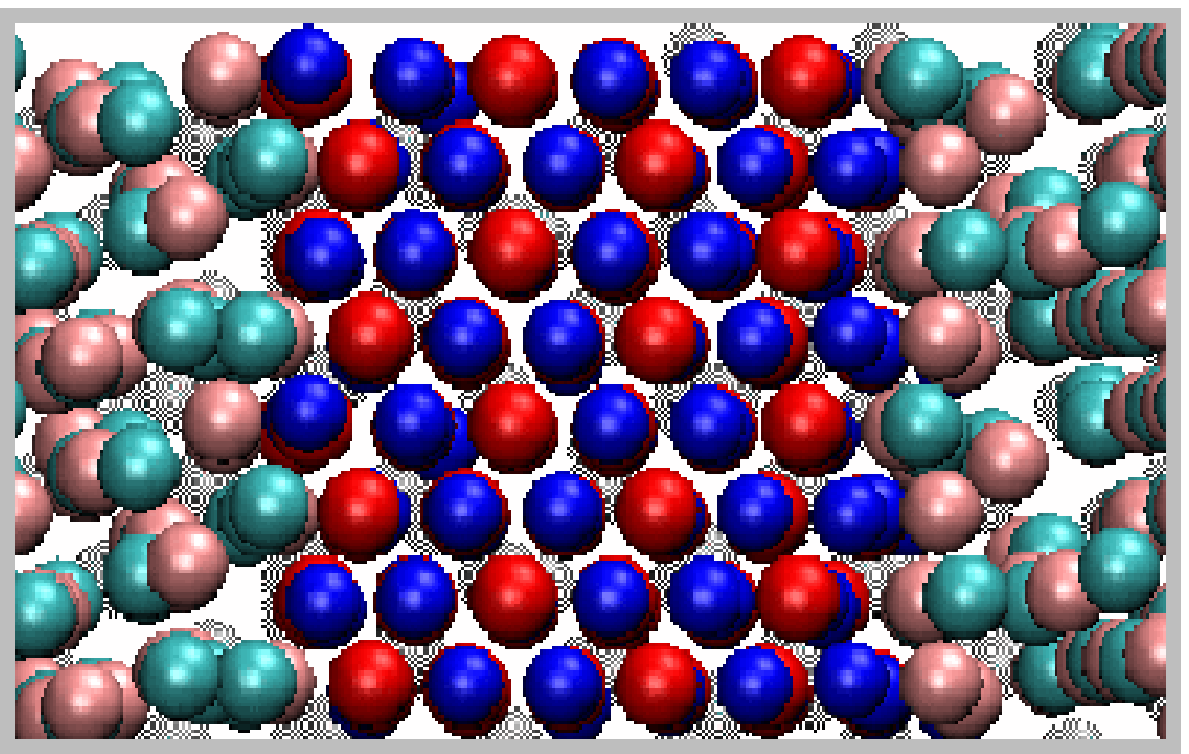} (d)}}
\caption[]
{\label{fig5} \noindent
Side and top views of mixed Li$_2$O/LiF films with grain boundaries on
Li metal surfaces.  (a)-(b): unstrained; (c)-(d): strained by 3~\AA.
Red and pink are O and F atoms.  Li is blue, cyan, and silver depending
on whether it starts out in Li$_2$O, LiF, or Li metal.  The Li metal
atoms underneath the film in the top view panels are depicted as transparent.
The circle indicates a disordered LiF region.
}
\end{figure}

Fig.~\ref{fig5} depicts mixed Li$_2$O/LiF films with grain boundaries on
Li metal surfaces.  When a 3~\AA\, strain is applied to this sytem (panels
(c)-(d)), cleavage of Li-F ionic bonds, rather than cleavage of Li$_2$O from
LiF, is observed.  Comparing the unstrained and strained configurations,
Fig.~\ref{fig5}a vs.~Fig.~\ref{fig5}c and Fig.~\ref{fig5}b
vs.~Fig.~\ref{fig5}d, reveals that the Li$_2$O region remains ordered
while LiF near the interface exhibits a disordered lattice structure.
Note that, by construction, the two LiF/Li$_2$O interfaces in the simulation
cell are different and respond to strain differently.  The grain boundary
atomic environment qualitatively resemble that in LiF, which does not
favor Li$^{(0)}$ insertion and has been discussed previously.  We have not
systematically examined inserting Li$^{(0)}$ here.

\begin{figure}
\centerline{\hbox{ \epsfxsize=3.00in \epsfbox{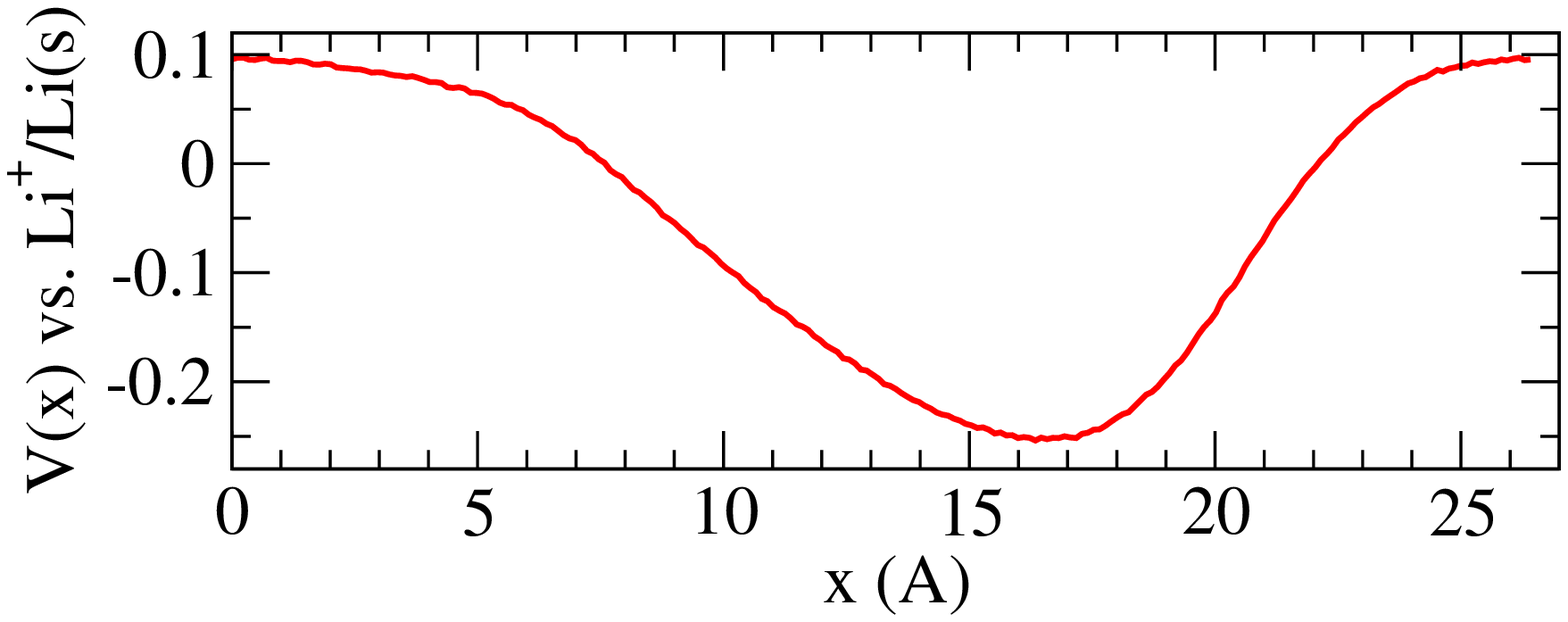} (a)}}
\centerline{\hbox{ \epsfxsize=3.00in \epsfbox{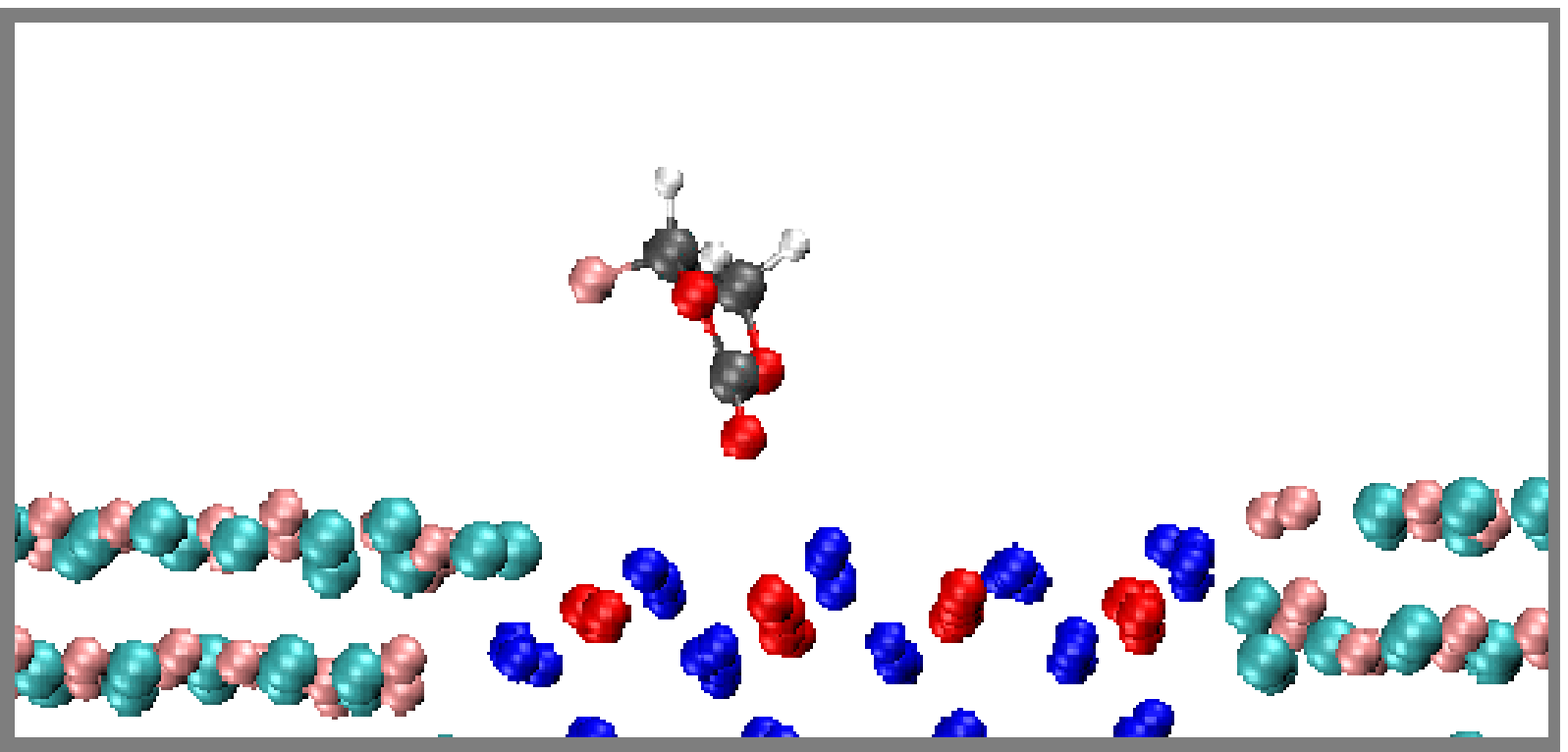} (b)}}
\centerline{\hbox{ \epsfxsize=3.00in \epsfbox{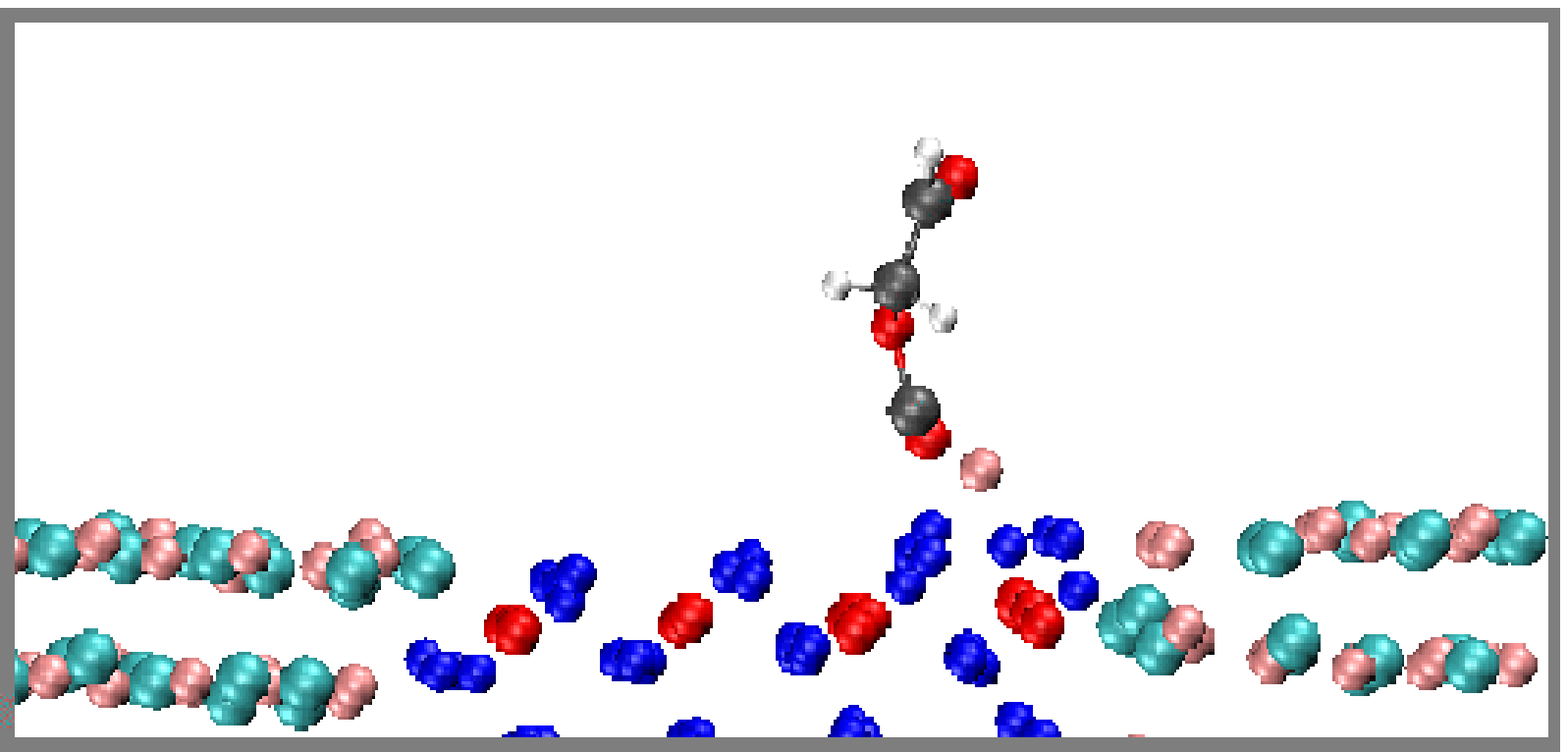} (c)}}
\caption[]
{\label{fig6} \noindent
(a) Mean ``local potential'' along the $x$-direction for
mixed LiF/Li$_2$O grain boundaries on Li(s) without FEC molecule;
(b) FEC remains intact at high potential grain boundary region;
(c) FEC decompose at low potential grain boundary region.
The Li anode in panels (b) and (c) are out of the frame.
}
\end{figure}

Since Li(s) is metallic, the thin-film-coated anodes studied in this work
each exhibits a single work function or $E_{\rm F}$ at all spatial points
inside the metallic region.  The unique work function for each model electrode
is obtained by averaging over the electrostatic potential $\phi(x,y,z)$ in the
$x$-$y$ plane at a $z$ position $z_o$ sufficiently deep into the vacuum
region, and using that as the zero energy reference.  However, an effective
local voltage $V(x)$ can be defined by averaging $\phi(x,y,z_o)$ over the
$y$-direction.  This function is useful for computing $e^-$ tunneling
probability at different $x$-positions.  Fig.~\ref{fig6}a depicts this
``local voltage'' for the unstrained mixed surface film (Fig.~\ref{fig5}a-b)
along the $x$-direction perpendicular to the grain boundaries.  $V(x)$ is
found to be inhomogeneous, with variations exceeding 0.25~V and its lowest
value at one of the LiF-Li$_2$O grain boundaries.

To illustrate the implications of $V(x)$ spatial inhomogeneity, we add
a fluoroethylene carbonate (FEC) molecule, a popular electrolyte
additive,\cite{fec1,fec2,fec3} at two $x$ positions.  The location at the
left-most grain boundary is at a higher voltage.  The FEC there has not
been electrochemically reduced after a 5~ps DFT-based molecular dynamics
trajectory at T=350~K.  At the end of the trajectory (Fig.~\ref{fig6}b),
this FEC has diffused to the almost middle of the Li$_2$O region, but
remains intact.  The rightmost grain boundary is at a substantially more
negative local potential, and is a ``hot spot'' for passivation breakdown.
Placing a FEC there and initiating molecular dynamics leads to FEC reductive
decomposition within 0.5~ps (Fig.~\ref{fig6}c).  Although anecdotal, this
evidence underscores the importance of spatial inhomogeneities to electrolyte
decomposition -- even in the absence of cracks or Li$^{(0)}$ in the grain
boundaries.

\subsection*{Films with Cracks}

\begin{figure}
\centerline{\hbox{ \epsfxsize=2.20in \epsfbox{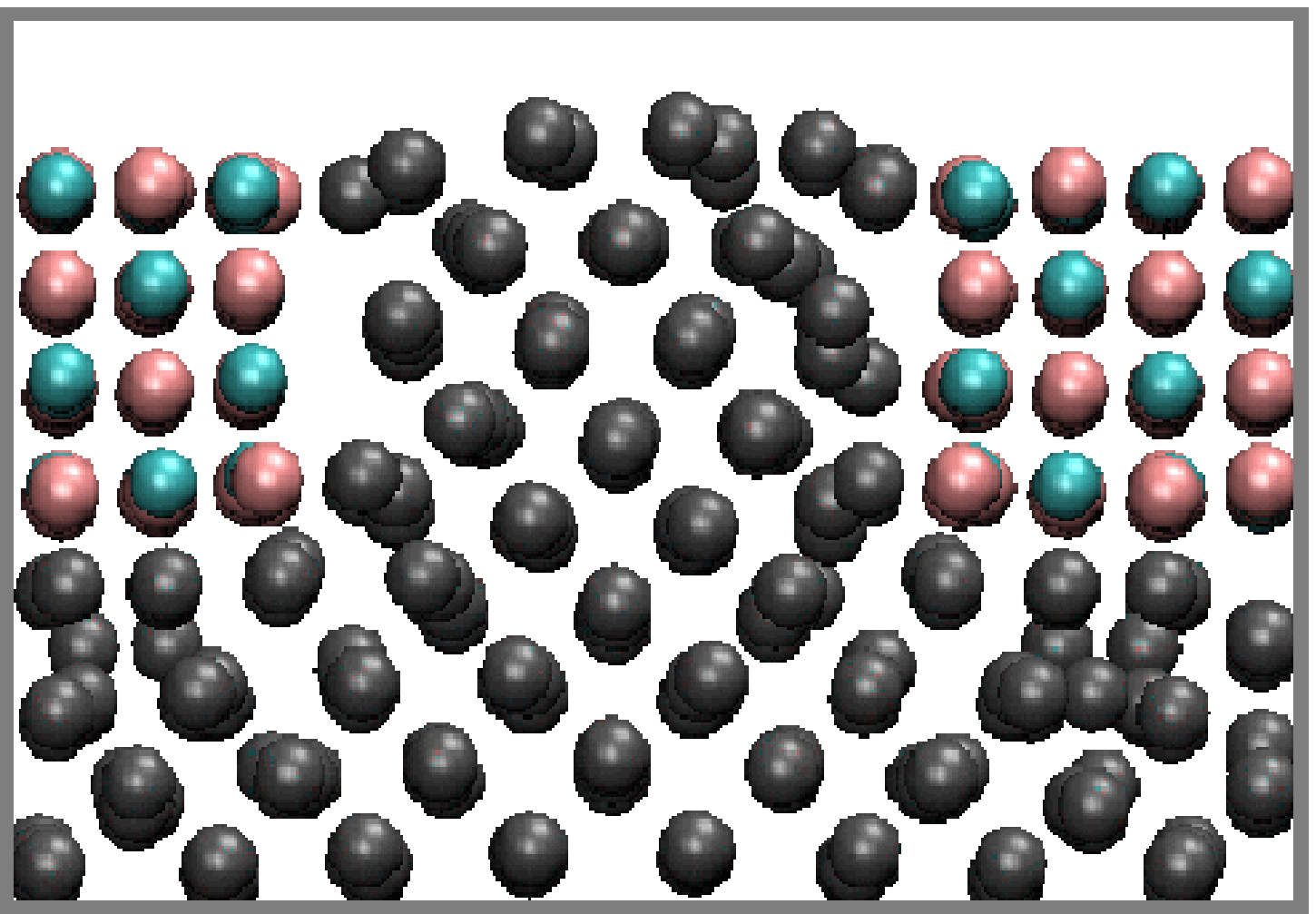} (a)  
		   \epsfxsize=2.20in \epsfbox{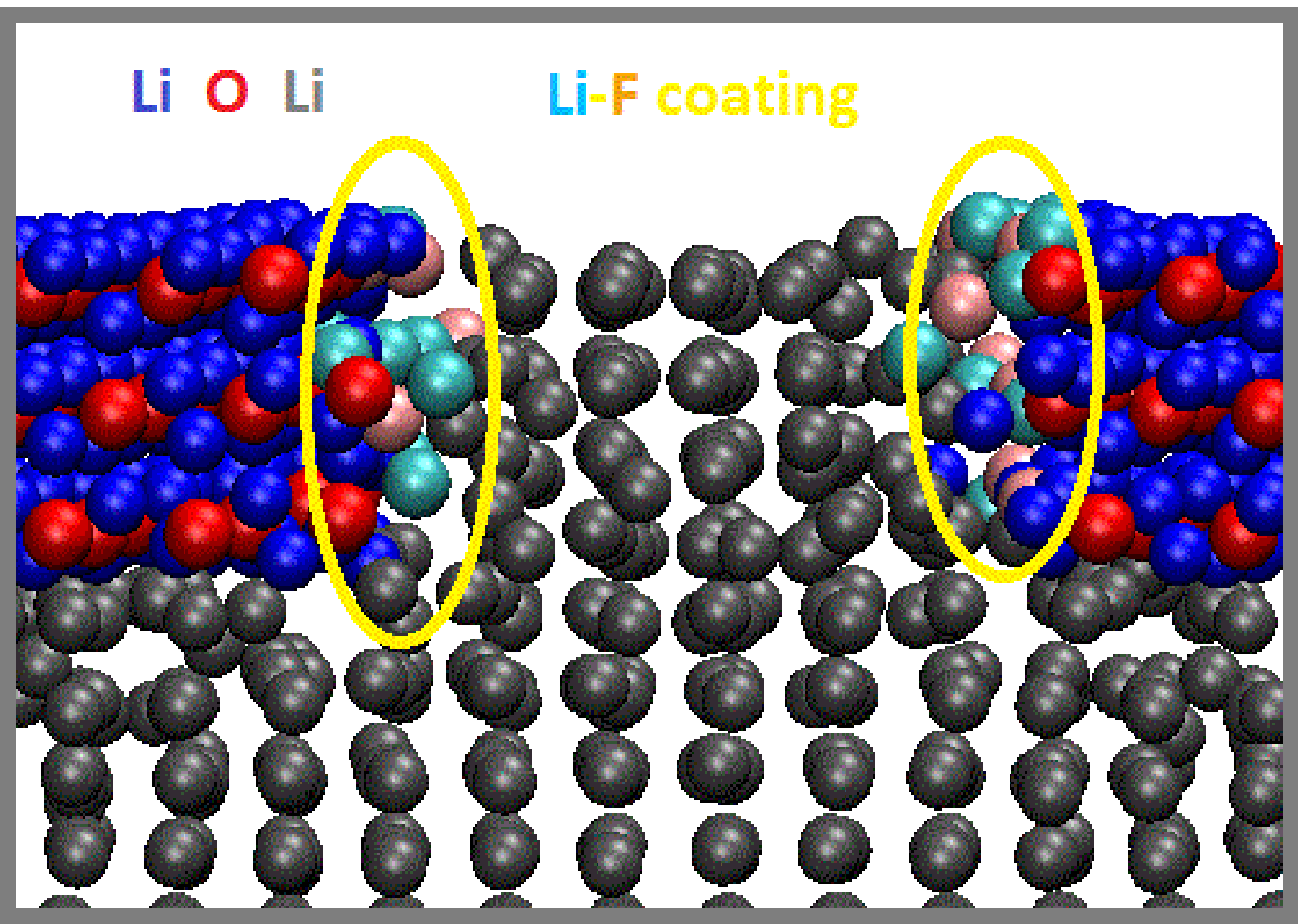} (b)}}
\caption[]
{\label{fig7} \noindent
(a)-(b) Side views of $\sim$12~\AA\, thick Li sheets (incipient ``lithium
filament'') inside a crack in a LiF film, and a Li$_2$O film decorated by
LiF dimers, respectively.  For color key, see Fig.~\ref{fig1}.
The ovas indicate LiF decoration on the oxide surfaces.
}
\end{figure}

Finally, in view of the atomic lengthscale crack developing in strained Li$_2$O
films, we also examine surface films with wider gaps or cracks to determine
whether Li nanosheets can grow there.  Fig.~\ref{fig1}f and~Fig.~\ref{fig7}a
depict $\sim$10~\AA\, thick Li$_2$O and LiF films, both with
$\sim$12~\AA\,-wide gaps, on Li(s) surfaces.  The computed overpotentials
needed for inserting body-center-cubic Li metal into these gaps (126 and 118
Li atoms) are +0.05 and $-0.10$~V, respectively.  It is therefore energetically
favorable to insert a Li(s) nanosheet into the Li$_2$O gap but not LiF.
Fig.~\ref{fig7}b depicts the Li$_2$O film with a gap decorated with LiF dimers
at a surface density of $\sim$2.9~nm$^{-2}$.  The configuration is optimized
before introducing 118~Li inside the gap in the cell. The computed overpotential
needed to insert that Li metal sheet is $-0.07$~V, suggesting that Li insertion
may be thermodynamically favorable at 0~V vs.~Li$^+$/Li(s).  While LiF
dimer-coated Li$_2$O yields an unfavorable monolayer Li adsorption energy
(Fig.~\ref{fig2}c), the F$^-$ anions at the surface can be readily absorbed
into the added Li metal nanosheet, negating that lithium-phobic condition.  A
thicker layer of LiF is apparently needed to impede Li intrusion.  These
predictions may have impact on future attempts at passivating boundaries or
cracks in SEI.

\section*{CONCLUSIONS}

In conclusion, this computational work illustrates the effects of atomic
lengthscale inhomogeneities in passivation films covering lithium anode
surfaces.  The lengthscale considered is too small to be conclusively imaged
using current experimental methods, but our predictions will help
motivate future experimental work.  Both LiF and Li$_2$O on Li metal
surfaces exhibit wide band gaps and block electrons if they are defect
free.\cite{yue}  But mobile Li$^{(0)}$ is found to reside in and diffuse
along Li$_2$O grain boundaries at $<0.25$~V computed overpotential if there is 
sufficient void space there.  Strain often accompanies Li$^+$ insertion
into anodes. Applying a 1.7\% strain lowers the computed overpotential needed
for Li$^{(0)}$ formation in the grain boundary to 0.1~V.  Furthermore,
there can be significant spatial fluctuations in the local potential.
As such, Li$^{(0)}$ species may be responsible
for through-SEI $e^-$ transport, initial passivation breakdown of surface
films, and slow increase of impedance at the interface as the battery ages.  
Upon further application of strain, subnanometer-sized particles of Li metal
can grow in atomic lengthscale gaps that develop at Li$_2$O grain boundaries,
forming ``incipient lithium filaments'' that may cause subsequent growth of
dendrites under adverse conditions.  These findings appear qualitatively
consistent with the fact that applying pressure, which reduces void spaces,
improves the performance of Li metal anodes.\cite{pressure} The negative
electron affinity material LiF is much more resistant to Li$^{(0)}$ insertion.
Our grain boundary models are meant to represent defected/amorphous
regions of SEI inorganic layers, the growth of which are kinetically
controlled.  Therefore we also illustrate the fundamental material and
surface difference between Li$_2$O and LiF by considering the energetics
of monolayer Li metal films on these surfaces.  The results suggest that
the difference between these materials in the SEI is likely to exist
independent of the specific model used.  This simple test can potentially be
used to examine the $e^-$-blocking ability of novel, artificial coating layers
and their defects.  In general, we postulate that additives
and new strategies need to mitigate passivation failures of Li anode
protection films at inhomogeneities, not just for defect-free films.

\section*{ACKNOWLEGEMENTS}
 
This work was performed, in part, at the Center for Integrated
Nanotechnologies, an Office of Science User Facility operated for the U.S.
Department of Energy (DOE) Office of Science. Sandia National Laboratories
is a multi-mission laboratory managed and operated by National Technology
and Engineering Solutions of Sandia, LLC., a wholly owned subsidiary of
Honeywell International, Inc., for the U.S. Department of Energy's National
Nuclear Security Administration under contract DE-NA-0003525.
KL, who performed the calculations, was supported by the Assistant Secretary
for Energy Efficiency and Renewable Energy, Office of Vehicle Technologies of
the U.S.~Department of Energy under Contract No. DE-AC02-05CH11231,
Subcontract No. 7060634 under the Advanced Batteries Materials Research (BMR)
Program.  KLJ, who provided the experimental motivation, was supported by
Nanostructures for Electrical Energy Storage (NEES), an Energy Frontier
Research Center funded by the U.S.~Department of Energy, Office of Science,
Office of Basic Energy Sciences under Award Number DESC0001160.

\section*{Supporting Information for Publication}
Supporting information is available free of charge on the ACS Publications
website at DOI: 

Details of models used; Li$^{(0)}$ diffusion along grain boundaries;
Li adsorption on planar surfaces; interfaces between crystalline
LiF/Li$_2$O and Li metal; Li$_2$O $\Sigma_5$ grain boundaries; model
with a thicker Li$_2$O surface film.

\end{document}